\documentclass[prd,showpacs,amssymb,amsmath,nofootinbib]{revtex4}
\usepackage{graphicx,xcolor,soul}

\newcommand{\e}{{\mathrm{e}}}

\newcommand{\CC}{{\mathbb C}} 
\newcommand{\RR}{{\mathbb R}} 
\newcommand{\NN}{{\mathbb N}}

\newcommand{\FF}{{\mathcal F}}

\newcommand{\Rc}{{\mathcal R}}

\DeclareMathOperator{\erfc}{erfc}
\DeclareMathOperator{\sgn}{sgn}
\DeclareMathOperator{\sinc}{sinc}
\DeclareMathOperator{\Tr}{Tr}

\newtheorem{theorem}{Theorem}[section]
\newtheorem{lemma}[theorem]{Lemma}

\begin{document}

\title{Probability Distributions for Quantum Stress Tensors 
\\ Measured in a Finite Time Interval}

\author{Christopher J. Fewster}
\email{chris.fewster@york.ac.uk}
\affiliation{Department of Mathematics, University of York, 
Heslington, York YO10 5DD,
United Kingdom}

\author{L. H. Ford}
\email{ford@cosmos.phy.tufts.edu}
\affiliation{Institute of Cosmology, Department of Physics and
  Astronomy, 
Tufts University, Medford, Massachusetts 02155, USA}

\begin{abstract}
A meaningful probability distribution for measurements of a quantum stress tensor operator can only be obtained if 
the operator is averaged in time or in spacetime. This averaging can be regarded as a description of the measurement process. Realistic measurements can be expected to begin
and end at finite times, which means that they are described by functions with compact
support, which we will also take to be smooth. Here we study the  probability distributions for stress tensor operators averaged
with such functions of time, in the vacuum state of a massless free field. Our primary aim is to understand the asymptotic form of the distribution
which describes the probability of large vacuum fluctuations. Our approach involves  asymptotic
estimates for the high moments of the distribution. These estimates in turn may be used
to obtain estimates for the asymptotic form of the probability distribution. Our results
show that averaging over a finite interval results in a probability distribution which falls
more slowly than for the case of Lorentzian averaging, and both fall more slowly than
exponentially. This  indicates that vacuum fluctuations effects can dominate over thermal
fluctuations in some circumstances.
\end{abstract}

\maketitle

\section{Introduction}
\label{sec:intro}

It is well known that the vacuum fluctuations of a linear quantum field operator are
associated with a Gaussian probability distribution.  More precisely, it is an averaged field operator that has a well defined probability distribution. Let 
$\varphi(t,{\bf x})$ be a field operator or a derivative of a field operator.
The averaging can be conducted along a timelike curve (or other
timelike submanifold), over a spatial volume, or over a spacetime volume (see, e.g.\ Ref.~\cite{Jaffe} for a discussion and references). Here we illustrate the case of temporal
averaging with a real-valued sampling function $f_\tau(t)$, with characteristic width $\tau$ which
satisfies
\begin{equation}
\int_{-\infty}^\infty f_\tau(t) \, dt =1 \,.
\label{eq:fnorm}
\end{equation} 
Define the averaged field operator by
\begin{equation}
\bar{\varphi} = \int_{-\infty}^\infty  \varphi(t,{\bf x}) \, f_\tau(t) \, dt \,.
\label{eq:t-ave}
\end{equation} 
This operator has finite moments in the vacuum state, $\mu_n = \langle \bar{\varphi}^n \rangle$,
which are those of Gaussian distribution; this may be summarized in the
formula
\begin{equation}
\sum_{n=0}^\infty \frac{(i\lambda)^n}{n!}\mu_n=
\langle \e^{i\lambda\bar{\varphi}}\rangle = \e^{-\frac{1}{2}\lambda^2\langle\bar{\varphi}^2\rangle}.
\end{equation}
Thus the odd moments vanish, and the even moments 
grow as $n!$ By the Hamburger moment theorem (See, for example, Ref.~\cite{Simon}), these 
moments uniquely define the probability distribution to be a Gaussian
\begin{equation}
P(\bar{\varphi})=\frac{1}{\sqrt{2\pi \sigma}}\,\exp\left(-\frac{\bar{\varphi}^2}{2\sigma}\right) \,.
\label{eq:pdf}
\end{equation}
Here the variance $\sigma=\langle\bar{\varphi}^2\rangle$ is determined by both the functional form and width of $f_\tau(t)$: crucially, $\sigma$ is finite provided the Fourier transform $\hat{f}(\omega)$ decays sufficiently fast as $\omega\to\infty$.\footnote{The same is true for averaging over spatial and spacetime volumes, but is not true for averaging over spacelike curves.} 
Some physical effects of time averaged quantum electric field fluctuations have been discussed
in Refs.~\cite{BDFS14,HF15}, in examples where the physical situation can define the sampling
function.

The probability distributions for quadratic operators, such as the energy density or other components
of the stress tensor,  were discussed in Refs.~\cite{FFR10,FFR12}. Here it
is an average of a normal ordered operator which has finite moments. In the case of the stress tensor
for conformal field theory in two spacetime dimensions with a Gaussian sampling function in either
space or time, the probability distribution is a shifted gamma distribution~\cite{FFR10} with the lower bound of its support given by the quantum inequality bound~\cite{FH05}. This includes the
case of the energy density of a massless scalar field, for which 
the quantum inequality bound was found by Flanagan~\cite{Flanagan97}.    

The corresponding problem for quadratic operators in four spacetime dimensions is more complicated. 
Here the sampling must be in time or spacetime, not space alone.
\footnote{See again Ref.~\cite{Jaffe};
more recent developments using microlocal analysis~\cite{BrunettiFredenhagen} permit a more
geometric framework that can be used to understand the reason spatial averaging is excluded. 
The basic issue is the ability to restrict distributions to submanifolds; such restrictions certainly
exist if the wavefront set of the distribution contains no covector that can annihilate all the tangent
vectors to the submanifold (see, e.g., the discussion of timelike curves in~\cite{AGWQI}). 
In the case of a (derivative of) a free field, the relevant wavefront set contains only null covectors, which 
cannot annihilate the tangent vectors to a timelike curve or a spacelike hypersurface of codimension $1$.
However, in the case of quadratic operators the relevant wavefront set contains not only null, but also timelike covectors, which can annihilate all the tangent vectors to a spacelike submanifold, so the 
restriction cannot be achieved by these means. There is a more heuristic argument for the inadequacy of spatial averaging alone.
A plane wave expansion of a quadratic operator will contain terms proportional to $\e^{i(\mathbf{k}+  \mathbf{k'})\cdot \mathbf{x}}$.
Modes for which $\mathbf{k} = - \mathbf{k'}$ will yield a divergent contribution to the moments which is not suppressed
by spatial averaging. This effect is the reason why there are no quantum inequalities for spatially averaged stress tensors
in four spacetime dimensions~\cite{FHR02}.  Similar problems affect averages along null curves~\cite{FR03}.   }
The case of Lorentzian sampling
in time of the stress tensor was treated in Ref.~\cite{FFR12}, where several features of the probability 
distribution were  inferred from the calculation of a finite set of moments.
(See also Ref.~\cite{FS14}, in which some combinatorial results
used in Ref.~\cite{FFR12} were proved.)
In this case, the moments grow 
as $(3n)!$, and neither the Hamburger moment condition nor the related Stieltjes condition is satisfied,
so  the moments are not guaranteed to define a unique 
probability distribution (and the available evidence suggests that they do not). However, the asymptotic behavior of the  probability distribution for large arguments
can be inferred from the moments. This allows us to estimate the probability of especially large fluctuations.
For example, let $u$ be the  Lorentzian average on a time scale $\tau$ of the electromagnetic energy density.
In units where $\hbar =c=1$, which will be used throughout this paper, the dimensionless measure of
the averaged energy density can be taken to be $x = u \tau^4$. The asymptotic form of the probability 
distribution for large $x$ is of the form
\begin{equation}
P(x) \sim x^{-2} \, \exp(-a\, x^{1/3}) \,,
\label{eq:Lor}
\end{equation}
where $a$ is a numerical constant of order one. The $x^{1/3}$ dependence in the exponential indicates
that large energy density fluctuations are more likely than one might have expected. Large thermal
fluctuations are suppressed by an exponential in energy, and hence can be dominated at larger 
energies by vacuum fluctuations. 

Quantum stress tensor fluctuations can have a variety of physical effects, and have been discussed
by several authors~\cite{WF01,Borgman,Stochastic,FW04,TF06,PRV09,FW07,CG97,WKF07,FMNWW10,WHFN11,LN05,Wu2007}.
These effects include radiation pressure fluctuations~\cite{WF01}, and passive quantum fluctuations
of gravity, where the gravitational field fluctuates in response to matter stress tensor fluctuations. Much
of the previous work on this topic has involved calculating mean squared effects using integrals of a
stress tensor correlation function. An exception is a study of black hole and Boltzmann brain nucleation
based on Eq.~\eqref{eq:Lor} in  Ref.~\cite{FFR12}. A deeper understanding of the probability distribution
will allow further investigation of the effects of large fluctuations.

The purpose of the present paper is to explore  the probability distributions associated with sampling
functions with compact support in time. These are functions which are strictly zero outside of a finite
interval, but can nonetheless be infinitely differentiable. Such functions seem to be the most appropriate
descriptions of measurements which begin and end at finite times, and avoid the long temporal tails
of functions like the Lorentzian. Some examples of such functions will be treated in Sec.~\ref{sec:compact}, and in particular it will be shown constructively that, for any $0<\alpha<1$, there are real-valued and nonnegative smooth compactly supported functions whose Fourier transforms are
of form $\exp(-\beta|\omega|^\alpha)$ up to a fractional error of order $|\omega|^{\alpha-1}$, where $\beta>0$. The parameter $\alpha$ controls the behavior of the test function near the switch-on and
switch-off regions; we also give a physical motivation for the use of such functions.
The asymptotic behavior of the moments of quadratic operators sampled with these functions will be
treated in Sec.~\ref{sec:large} and shown to have a dominant contribution that grows like
$\Gamma((np+2)/\alpha)$, where $p$ depends on the operator in question ($p=3$ for the
energy density). The resulting asymptotic form for high moments will be used in
Sec.~\ref{sec:tail} to infer the tail of the probability distribution for large arguments, and hence to
estimate the probability for large fluctuations. The tail takes the form $P(x)\sim c_0 x^{b}\exp(-a x^{\alpha/p})$ for certain parameters $c_0$, $a$ and $b$, and it will be shown that the probability of
large fluctuations can be much greater than that for the case of Lorentzian averaged operators. Our results and their possible applications will be discussed in Sec.~\ref{sec:sum}. Three appendices contain
technical details on various estimates used in the text.

\section{Compactly Supported Sampling Functions}
\label{sec:compact}

\subsection{General results and constructions}

The sampling functions studied in this paper will be smooth and compactly supported
nonnegative functions, which model a sampling period of finite duration. At the outset
it is worth recalling some facts concerning the Fourier transforms of such functions. Let
$f$ be smooth and compactly supported. Then its transform
\begin{equation}\label{eq:Fourier}
\hat{f}(\omega) =\int_{-\infty}^\infty dt \,\e^{-i\omega t} f(t)
\end{equation}
is clearly bounded, $|\hat{f}(\omega)|\le \int_{-\infty}^\infty dt |f(t)|$, and as this conclusion
applies equally to the derivatives of $f$, we deduce that, for any $n\in\NN$,
\begin{equation}
|\hat{f}(\omega)| =  \frac{|\hat{f}(\omega)+(-1)^n\widehat{f^{(2n)}}(\omega)|}{1+\omega^{2n}}
\le \frac{C}{1+\omega^{2n}}
\end{equation}
for some constant $C$ depending on $f$ and $n$. This is the well-known statement that
the transform decays faster than any inverse power. On the other hand, the transform cannot
decay too rapidly: if there are positive
constants $A$ and $\alpha$ such that $|\hat{f}(\omega)|\le A \e^{-\alpha |\omega|}$ for all $\omega$, 
then the inversion formula
\begin{equation}
f(t)=\int_{-\infty}^\infty \frac{d\omega}{2\pi}\, \e^{i\omega t} \hat{f}(\omega) 
\end{equation}
provides an analytic extension of $f$ to the strip $|\Im z|<\alpha$ of the complex plane. As $f$ vanishes with all its derivatives outside its support, it must therefore vanish identically by uniqueness
of analytic continuation.  However, Ingham~\cite{Ingham} 
has shown that there are functions of compact support with transforms obeying
$\hat{f}(\omega)=O(\exp(-|\omega|\eta(|\omega|)))$ as $|\omega|\to\infty$ for 
any positive function $\eta$ decreasing monotonically to zero as $\omega\to\infty$ and such that
\begin{equation}
\int_1^\infty d\omega\,\frac{\eta(\omega)}{\omega} <\infty.
\end{equation}
In particular, Ingham's result applies to any function $\eta(\omega)=\beta \omega^{\alpha-1}$ 
for $0<\alpha<1$ and $\beta>0$ establishing the existence of smooth compactly supported functions $f$
with $\hat{f}(\omega)=O(\e^{-\beta |\omega|^\alpha})$. 
At this stage, we remark that our usage of the `big-O'  notation will
coincide with that in Ref.~\cite{deBruijn}: to write that $g(s)=O(h(s))$
on some set $S$ means that there exists a constant $A>0$ such that
$|g(s)|\le A|h(s)|$ for all $s\in S$ -- in fact we will only ever use this
notation in situations where $h(s)$ is nonnegative, so one has $|g(s)|\le Ah(s)$. To write that `$g(s)=O(h(s))$ as $s\to\infty$' indicates
that there exists $s_0>0$ such that $g(s)=O(h(s))$ on $[s_0,\infty)$. 

Ingham's construction defines the desired test function $f$ in terms of its Fourier transform, which is a carefully arranged infinite product of $\sinc$ functions. The precise behavior of $f(t)$ (which can be regarded as an infinite convolution of top hat functions) is not investigated.  Instead, we give
a different construction that is slightly more explicit. To start, let $\varphi\in C^\infty(\RR)$ be a smooth nonnegative and integrable function so that $\varphi(t)=0$ for $t\le 0$ and $\varphi(t)>0$ for $t>0$, with $\varphi(t)\to 0$ as $t\to +\infty$. Choosing $\delta>0$, we
may construct a smooth function with support $[-\delta,\delta]$ by setting
\begin{equation}
H(t) = \varphi(t+\delta)\varphi(\delta-t).
\end{equation}
The choice of $\delta$ can be used to tune the shape of $H$. 
For instance, if $\varphi$ has its first local maximum at $\delta$, then
elementary differentiation shows easily that $H$ has a local maximum at $t=0$. On the other hand, we can achieve a broader peak by taking $\delta$ slightly beyond
the maximum of $\varphi$, at the expense of making $t=0$ into a local minimum of $H$; this is what we will do in the specific examples below.  

As $\varphi$ is smooth and integrable, it has a Fourier transform, which decays rapidly.
Hence the convolution theorem applies,
giving
\begin{equation}\label{eq:convolution}
\hat{H}(\omega) =\e^{i\delta\omega} \int_{-\infty}^\infty \frac{d\omega'}{2\pi} \hat{\varphi}(\omega-\omega')\hat{\varphi}(-\omega')
\e^{-2i\delta\omega'}
\end{equation}
where we have used elementary results on Fourier transforms of shifts and reflections, i.e., $\FF[\varphi(t+\delta)](\omega)=  \e^{i\omega\delta}\hat{\varphi}(\omega)$ 
and $\FF[\varphi(\delta-t)](\omega) =   \e^{-i\omega\delta}\hat{\varphi}(-\omega)$.
For $\omega>0$, it is convenient to rewrite this expression as 
\begin{equation} \label{eq:simpleconv}
\hat{H}(\omega) =2\Re \left( \e^{i\delta\omega} I(\omega)\right), \qquad
I(\omega)= 
 \int_{-\omega/2}^\infty \frac{d\omega'}{2\pi} \hat{\varphi}(\omega+\omega')\hat{\varphi}(\omega')
\e^{2i\delta\omega'}   
\end{equation}
which may be proved by splitting the integration region in Eq.~\eqref{eq:convolution} into $(-\infty,\omega/2)$ and $(\omega/2,\infty)$
and making the changes of variable $\omega'\to -\omega'$ and
$\omega'\to\omega'-\omega$ in the resulting integrals, thus
yielding  
\begin{equation}
\e^{i\delta\omega} 
 \int_{-\omega/2}^\infty \frac{d\omega'}{2\pi} \hat{\varphi}(\omega+\omega')\hat{\varphi}(\omega')
\e^{2i\delta\omega'} 
\quad\text{and}\quad
\e^{-i\delta\omega} 
 \int_{-\omega/2}^\infty \frac{d\omega'}{2\pi} \hat{\varphi}(-\omega') \hat{\varphi}(-\omega-\omega')
\e^{-2i\delta\omega'},
\end{equation}
which are seen to be mutually complex conjugate on recalling that
$\hat{\varphi}(-\omega)=\hat{\varphi}^*(\omega)$ because
$\varphi$ is real-valued. As $H$ is real and even, the same is true of
$\hat{H}$, and therefore Eq.~\eqref{eq:simpleconv} can be modified to
hold for all $\omega$ by replacing all occurrences of $\omega$ on the 
right-hand side by $|\omega|$. 

Our interest is in the asymptotic behavior of $\hat{H}$ as $|\omega|\to\infty$. The specific examples chosen for 
$\varphi$ below have the property that the derivative $\hat{\varphi}'$ decays
more rapidly than $\hat{\varphi}$, which itself decays rapidly. Under these circumstances, the factors of 
$\hat{\varphi}(\omega+\omega')$ in the definition of $I(\omega)$ may be taken as approximately constant over the effective region of integration,
leading to the approximation
\begin{equation}
\hat{H}(\omega) \simeq  2\Re \left(\e^{i\delta\omega}\hat{\varphi}(\omega) 
 \int_{-\omega/2}^\infty \frac{d\omega'}{2\pi}  \hat{\varphi}(\omega')
\e^{2i\delta\omega'}  \right) \,.
\end{equation}
(We use $\simeq$ for an uncontrolled or heuristic approximation, 
reserving the symbol $\sim$ for a rigorously established asymptotic equivalence~\cite{deBruijn}.)
Furthermore, the rapid decay of $\hat{\varphi}$ allows us to extend the
integration range to the full real line, so that the integral becomes $\varphi(2\delta)$. Overall, we have
the approximation
\begin{equation}
\hat{H}(\omega)\simeq 2\varphi(2\delta) \Re \left( \hat{\varphi}(\omega)\e^{i\delta\omega}\right)
\end{equation}
provided the errors incurred can be suitably controlled. This is achieved in
Appendix~\ref{appx:int-ests}, with the following result (a particular application of Lemma~\ref{lem:int-ests}).
\begin{theorem}\label{thm:I-ests}
 Let $\beta>0$ and $0<\alpha<1$. Suppose the function $\varphi$
has transform obeying $\hat{\varphi}(\omega)=O(\e^{-\beta|\omega|^\alpha})$ and
$\hat{\varphi}'(\omega)=O(|\omega|^{\alpha-1}\e^{-\beta|\omega|^\alpha})$ for
all $\omega\in\RR$, and $\hat{\varphi}''(\omega)=O(|\omega|^{\alpha-2}\e^{-\beta|\omega|^\alpha})$ 
for $|\omega|>\omega_0$, for some $\omega_0\ge 0$. Then the integral $I(\omega)$
defined in Eq.~\eqref{eq:simpleconv} obeys
\begin{equation}\label{eq:I-omega}
I(\omega) = \varphi(2\delta) \hat{\varphi}(\omega) + O(|\omega|^{\alpha-1}\e^{-\beta|\omega|^\alpha})
\end{equation}
for all $\omega$, and 
\begin{equation}\label{eq:Iprime-omega}
I'(\omega) = \varphi(2\delta) \hat{\varphi}'(\omega)  + O(|\omega|^{\alpha-2}\e^{-\beta|\omega|^\alpha})
\end{equation}
for $|\omega|>\omega_0$. 
\end{theorem}
Consequently, under the hypotheses of the theorem, we have
\begin{equation}\label{eq:simpleapprox}
\hat{H}(\omega)= 2 \varphi(2\delta) \Re \left( \hat{\varphi}(\omega)\e^{i\delta\omega}\right)
+O(|\omega|^{\alpha-1}\e^{-\beta|\omega|^\alpha}) 
\end{equation}
for all $\omega$,
which shows in particular that $\hat{H}(\omega)=O(\e^{-\beta|\omega|^\alpha})$.
The derivative of $\hat{H}$ will be studied later, and we will also give specific examples of
functions $\varphi$ that meet the hypotheses of Theorem~\ref{thm:I-ests}, after giving
two further refinements of our construction.

First,  it will be convenient to work with a function that has a nonnegative transform. 
This is achieved by the expedient of taking the convolution of $H$ with itself, defining a smooth function
\begin{equation}\label{eq:Kdef}
K(t) = \int_{-\infty}^\infty dt'\,H(t-t')H(t')
\end{equation}
which has support $[-2\delta,2\delta]$ and transform $\hat{K}(\omega)=\hat{H}(\omega)^2$ obeying
\begin{equation}
\hat{K}(\omega) = 2\varphi(2\delta)^2\left(|\hat{\varphi}(\omega)|^2 +\Re \left( 
\hat{\varphi}(\omega)^2\e^{2i\delta\omega} \right)\right) + O(|\omega|^{\alpha-1}\e^{-2\beta|\omega|^\alpha}).
\end{equation}  
The behavior of $K(t)$ near the endpoints $t=\pm2\delta$ of the 
support is worth studying. Owing to the support of $H$, the integration range in Eq.~\eqref{eq:Kdef} may be replaced by $[-\delta,\delta]$ and, 
changing variables to $s=t'+\delta$,
we may calculate
\begin{equation}\label{eq:Kest}
K(-2\delta+\epsilon) = \int_{0}^{2\delta} ds\,
\varphi(\epsilon-s)\varphi(s) \varphi(2\delta-s)\varphi(2\delta+s-\epsilon)
\simeq  \varphi(2\delta)^2 \int_{0}^{\epsilon} ds\,
\varphi(\epsilon-s)\varphi(s)
\end{equation}
if $\epsilon\ll \delta$, using the support of the $\varphi(\epsilon-s)$ factor to restrict the integration range. The last integral may be recognised as a half-line convolution and is therefore the inverse Laplace transform of $\tilde{\varphi}(p)^2$. 

The second refinement arises because $\hat{K}$ has oscillations of magnitude comparable to its local mean value. It will be useful to damp these by a further trick: exploiting again the fact that the derivative
of $\hat{\varphi}$ falls off more rapidly than it does, the function 
\begin{equation}\label{eq:Ldef}
\hat{L}(\omega) = \hat{K}(\omega) + \frac{1}{2}\left(\hat{K}(\omega+\pi/(2\delta))+\hat{K}(\omega-\pi/(2\delta))\right)
\end{equation}
has asymptotic form
\begin{equation}
\hat{L}(\omega) =  4\varphi(2\delta)^2 |\hat{\varphi}(\omega)|^2 +O(|\omega|^{\alpha-1}\e^{-2\beta|\omega|^\alpha})
\end{equation}
valid for $|\omega|>\omega_0$ for any $\omega_0>\pi/(2\delta)$. This is proved by using
the estimates on $\hat{\varphi}$ and $\hat{\varphi}'$ in the hypotheses of Theorem~\ref{thm:I-ests},
the fact that $|\omega-\pi/(2\delta)|^{\gamma}\e^{-2\beta|\omega-\pi/(2\delta)|^\alpha}= 
O(|\omega|^{\alpha-1}\e^{-2\beta|\omega|^\alpha})$ on $|\omega|>\omega_0$,  and the following result. 
\begin{lemma} \label{lem:osc}
If $F(\omega)=f(\omega)\e^{2i\delta\omega}$ where $\delta>0$ is fixed and $f$ is differentiable, then
\begin{equation} 
F(\omega)+\frac{1}{2}\left(F(\omega+\pi/(2\delta))+F(\omega-\pi/(2\delta))\right) = 
O\left(\sup_{|\zeta-\omega|<\pi/(2\delta)} |f'(\zeta)|\right) \,.
\end{equation}
\end{lemma}
\emph{Proof} Elementary calculation shows that the right-hand side is $\frac{1}{2}\e^{2i\delta\omega}\left[f(\omega)-f(\omega-\pi/(2\delta)) +  f(\omega)-f(\omega+\pi/(2\delta))\right]$ and we then use the mean value theorem. $\square$

The function $\hat{L}$ may be recognised as the transform of 
\begin{equation}
L(t) = (1+\cos(\pi t/(2\delta))) K(t) = 
2\cos^2(\pi t/(4\delta))K(t), 
\end{equation}
which still has support $[-2\delta,2\delta]$.   Moreover, the
cosine factor executes a single period of oscillation on the support of $K$, and so does not
disturb the general `bump function' shape.

We also want to control the derivative $\hat{L}'(\omega)$. 
Recalling that $\hat{H}(\omega)=2\Re(\e^{i\omega\delta} I(\omega))$, it is easily seen that
\begin{equation}
\hat{H}'(\omega) = 2\delta\Re (i\e^{i\omega\delta}I(\omega)) + 
O(|I'(\omega)|)
\end{equation}
and therefore $\hat{K}=\hat{H}^2$ has derivative
\begin{align}
\hat{K}'(\omega) &= 2i\delta\left(\e^{i\omega\delta}I(\omega) + \e^{-i\omega\delta} I^*(\omega) \right)
\left(\e^{i\omega\delta}I(\omega) - \e^{-i\omega\delta} {I^*(\omega)}\right) + O(|I(\omega)I'(\omega)|) \\
&= 4\delta\Re(i \e^{2i\omega\delta}I(\omega)^2) + O(|I(\omega)I'(\omega)|) .
\end{align}
Substituting this expression into the derivative of  \eqref{eq:Ldef}
and applying  Lemma~\ref{lem:osc}, 
we find
\begin{equation}
\hat{L}'(\omega) = O\left( \sup_{|\zeta-\omega|<\pi/(2\delta)}|I(\zeta)I'(\zeta)|\right)=  
 O(|\omega|^{\alpha-1}\e^{-2\beta|\omega|^\alpha})
\end{equation}
valid for $|\omega|>\omega_0>\pi/(2\delta)$. 

\subsection{Specific examples}
\label{sec:examples}

A class of specific examples can be obtained as follows. 
We choose $\varphi$ to be the inverse Laplace transform of
$\tilde{\varphi}(p) = \exp(-(p\tau)^\alpha)$ where $\tau$ and
$\alpha$ are positive constants with $0<\alpha<1$. The role of $\tau$ is to set a timescale. As usual, 
the Laplace transform is defined by $\tilde{\varphi}(p) = \int_0^\infty dt\,
\varphi(t)\e^{-pt}$, and the branch cut for fractional
powers is taken to lie along the negative real axis.
Functions of this form have a number of interesting properties, some of
which will be discussed in Sec.~\ref{sec:Fox}. 
Given this definition, it is easily seen that $\varphi$ is supported in the positive half-line: 
if $t<0$ one may, in the Laplace inversion integral, complete the contour in the right-hand half-plane, in which $\tilde{\varphi}$ is analytic, and conclude that $\varphi(t)=0$. Consequently,
the Fourier transform $\hat{\varphi}$ of $\varphi$ 
extends to an analytic function in the lower half-plane and may be found as the boundary value 
\begin{equation}
\hat{\varphi}(\omega) =  \lim_{\epsilon\to 0+}\hat{\varphi}(\omega-i\epsilon)  =
 \lim_{\epsilon\to 0+} \tilde{\varphi}( \epsilon + i\omega) \,.
\end{equation}

This gives
\begin{equation}
\hat{\varphi}(\omega) = \exp\left(- \e^{i\sgn(\omega)\pi\alpha/2} |\omega\tau|^\alpha\right),
\end{equation}
from which we may read off that
\begin{equation}
|\hat{\varphi}(\omega)| = \exp\left(- \beta|\omega |^\alpha \right), \qquad \beta=\tau^\alpha\cos\frac{\pi\alpha}{2}.
\end{equation}
The derivatives of $\hat{\varphi}$ exist except at $\omega=0$, and obey 
\begin{equation} 
|\hat{\varphi}'(\omega)| = \frac{\alpha\tau^\alpha|\hat{\varphi}(\omega)|}{|\omega |^{1-\alpha}},\qquad
|\hat{\varphi}''(\omega)| \le \left(\frac{\alpha^2\tau^{2\alpha} }{|\omega |^{2(1-\alpha)}} +
\frac{\alpha(1-\alpha)\tau^\alpha}{|\omega |^{2-\alpha}}\right)
|\hat{\varphi}(\omega)|  \,.
\end{equation} 
In particular, $\hat{\varphi}''(\omega)=O(|\hat{\varphi}(\omega)|
|\omega|^{\alpha-2})$ for $\omega\in[\omega_0,\infty)$ for any
$\omega_0>0$. The hypotheses of Theorem~\ref{thm:I-ests} are therefore satisfied
and the construction above leads to a smooth, even and nonnegative function $L:\RR\to[0,\infty)$ with compact support in $[-2\delta,2\delta]$ and with Fourier transform that is 
analytic, even and nonnegative and obeys  
the estimates 
\begin{equation}\label{eq:Lprops1}
\hat{L}(\omega) =  4\varphi(2\delta)^2 \e^{-2\beta|\omega|^\alpha} +
O\left(\frac{\e^{-2\beta|\omega|^\alpha}}{|\omega|^{1-\alpha}}\right) ,
\end{equation}
for $\omega\neq 0$, and 
\begin{equation}\label{eq:Lprops2}
\hat{L}'(\omega) = O\left(\frac{\e^{-2\beta|\omega|^\alpha}}{|\omega|^{1-\alpha}}\right)
\end{equation}
valid for $\omega>\omega_0$ with $\omega_0>\pi/(2\delta)$. 
If, as suggested above, $\delta$ is chosen to be the first local maximum of $\varphi$,
then $\delta\propto\tau$, and (in any case) can clearly be arranged so that the width of the support takes
any desired value.

Meanwhile, the overall integral of $L$ is $\hat{L}(0)=\hat{K}(0)+\hat{K}(\pi/(2\delta)) =
\hat{H}(0)^2+ \hat{H}(\pi/(2\delta))^2$, and we may divide by this quantity to obtain a function with
unit integral all the previously stated properties and additionally having a unit integral.
Note that the coefficient of $\e^{-2\beta|\omega|^\alpha}$ in the normalized version of Eq.~\eqref{eq:Lprops1} is then 
$4\varphi(2\delta)^2/(\hat{H}(0)^2+ \hat{H}(\pi/(2\delta))^2)$. 

The approximation \eqref{eq:Kest} gives the behavior in $t$-space
near the end-points, noting that $\tilde{\varphi}(p)^2=\tilde{\varphi}(2^{1/\alpha}p)$ has inverse transform $2^{-1/\alpha}\varphi(t/2^{1/\alpha})$, 
\begin{equation}
K(-2\delta+\epsilon)\simeq \frac{\varphi(2\delta)^2}{2^{1/\alpha}}
\varphi(\epsilon/2^{1/\alpha}), \qquad
L(-2\delta+\epsilon)\simeq \frac{\pi^2}{8\delta^2}\epsilon^2K(-2\delta+\epsilon)
\end{equation}
for $\epsilon\ll\delta$ (here $L$ has not been normalized to unit integral).

\begin{figure}
 \centering
 \includegraphics[scale=0.8]{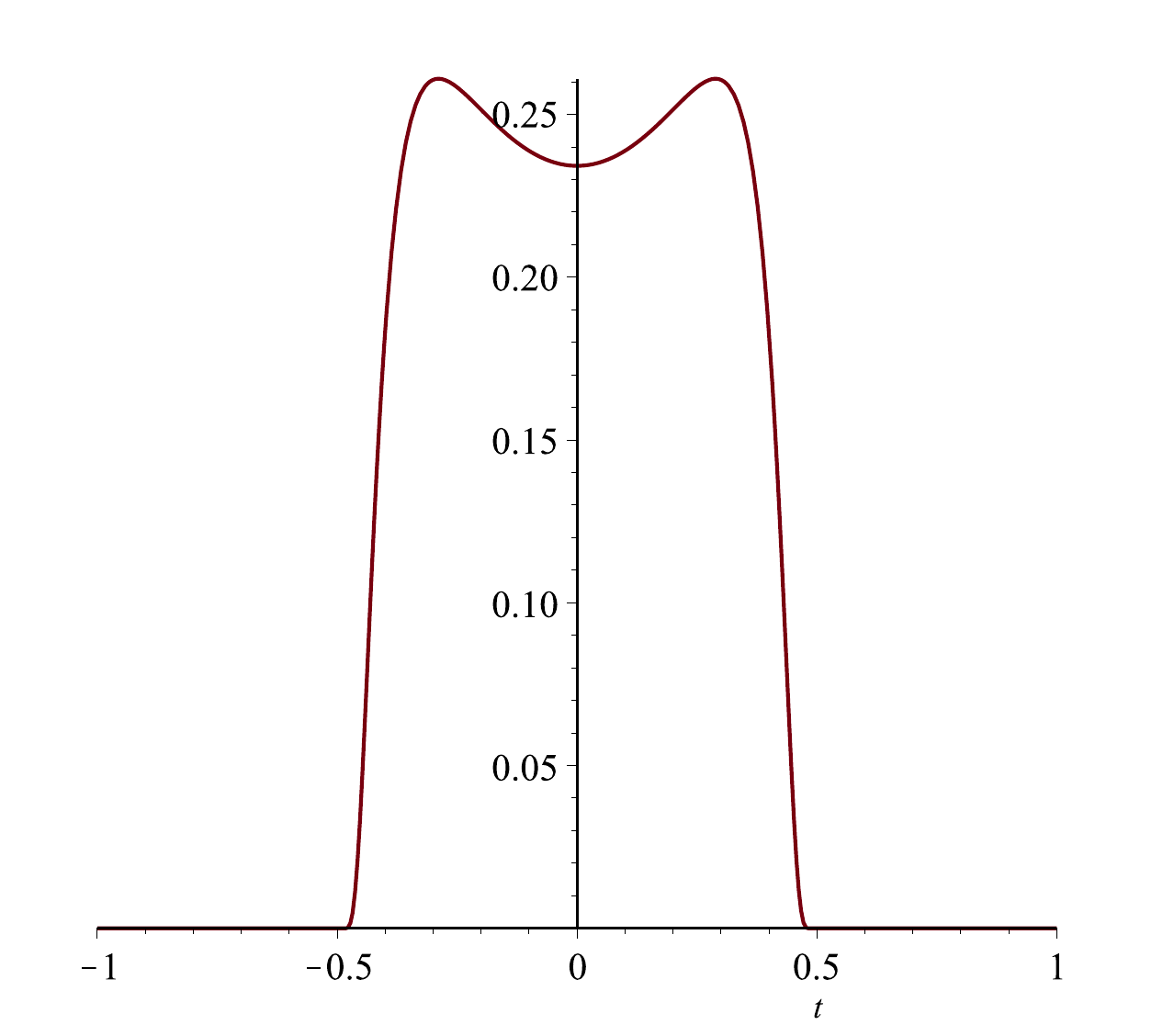}
 \caption{A plot of $H(t)$ for the case $\alpha = \delta=1/2$. } 
 \label{fig:H}
 \end{figure}

\begin{figure}
 \centering
 \includegraphics[scale=0.8]{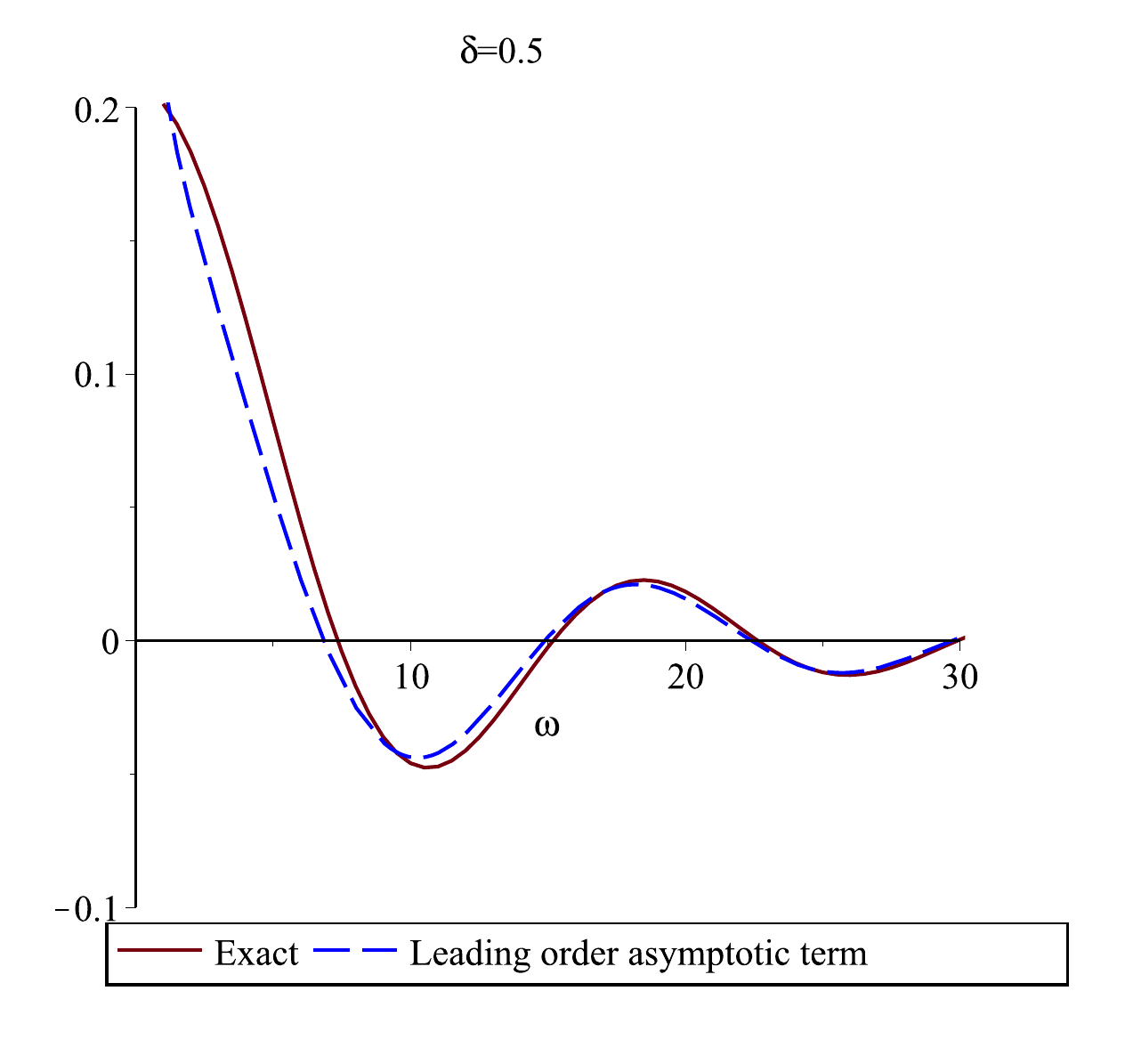}
 \caption{A plot of $\hat{H}(\omega)$ and its leading order approximation for the case $\alpha = \delta=1/2$. } 
 \label{fig:Hhat}
 \end{figure}

\begin{figure}
 \centering
 \includegraphics[scale=0.8]{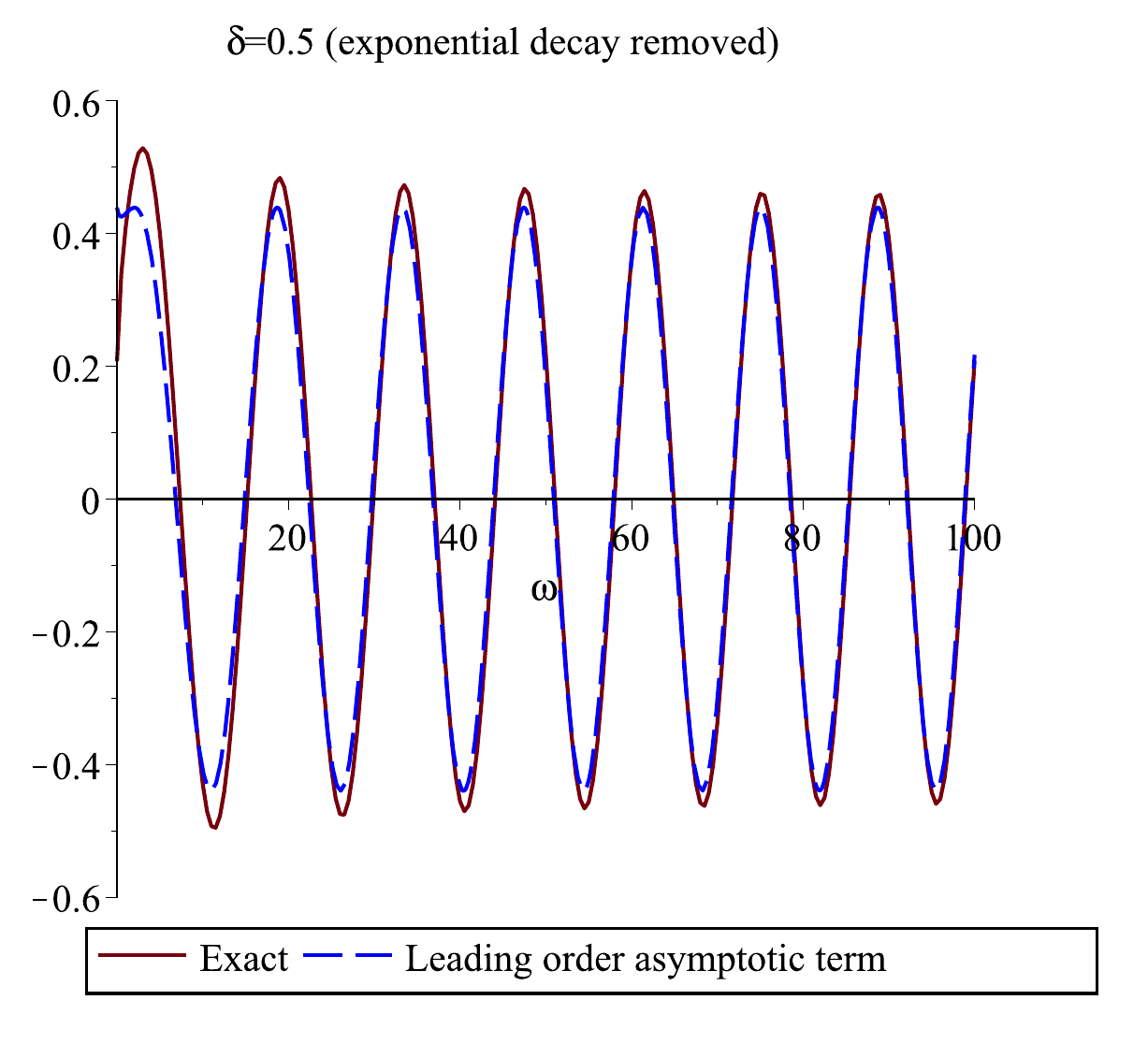}
 \caption{A plot of $\e^{\omega/2}\,\hat{H}(\omega)$ and its leading order approximation for the case $\alpha = \delta=1/2$. 
 Here the exponential decay has been removed to leave an envelope of approximately constant amplitude.} 
 \label{fig:Hhat-deexp}
 \end{figure}

\begin{figure}
 \centering
 \includegraphics[scale=0.8]{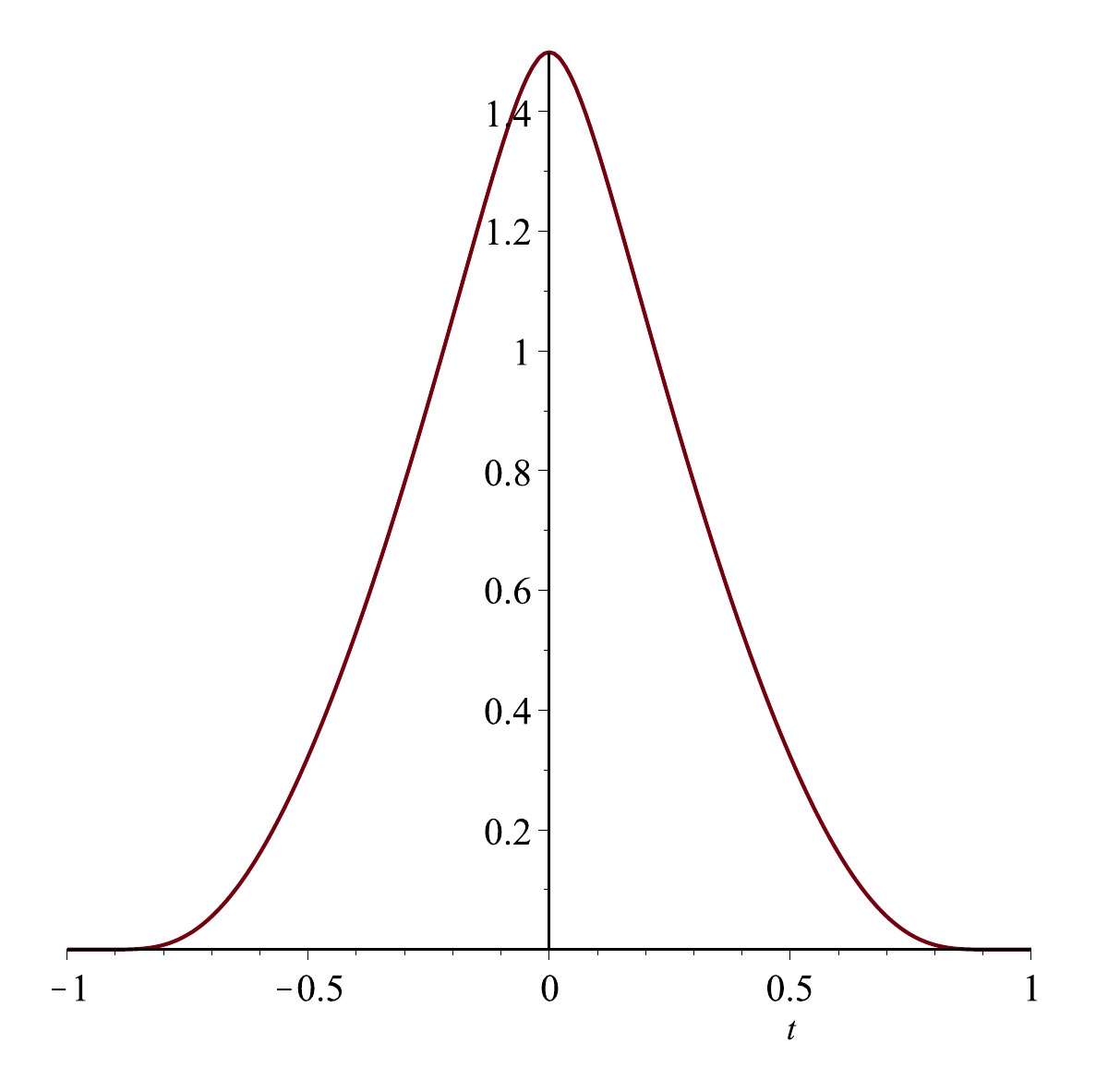}
 \caption{A plot of $L(t)$ 
 for the case $\alpha = \delta=1/2$. } 
 \label{fig:L-norm}
 \end{figure}

\begin{figure}
 \centering
 \includegraphics[scale=0.8]{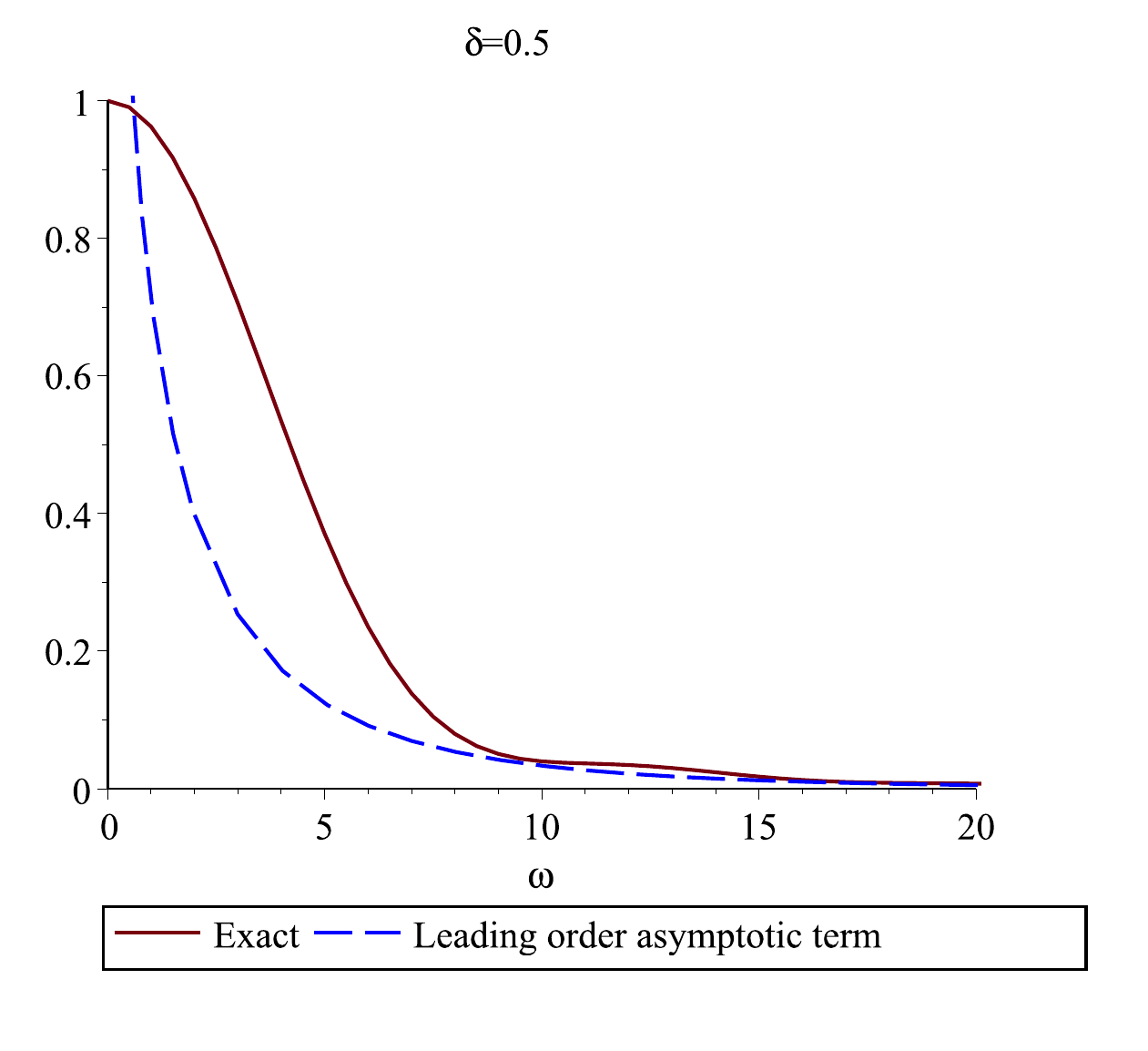}
 \caption{A plot of $\hat{L}(\omega)$ and its leading order approximation for the case $\alpha = \delta=1/2$. 
 Here $\hat{L}$ has been normalized so that $L$ has unit integral. The asymptotic approximation becomes good for $|\omega|\gtrsim 10$.} 
 \label{fig:Lhat-norm}
 \end{figure}

\begin{figure}
 \centering
 \includegraphics[scale=0.8]{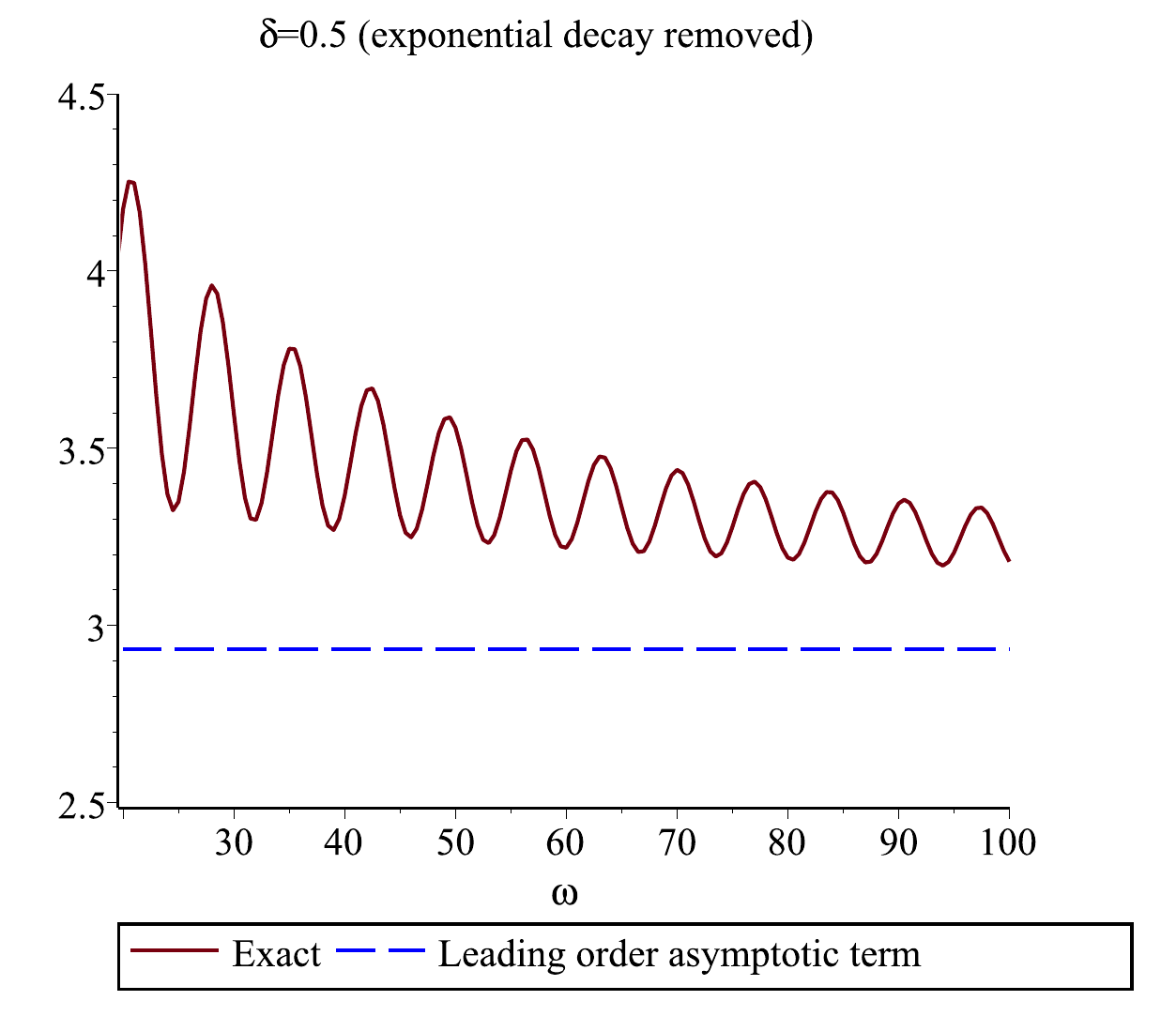}
 \caption{A plot of $\e^{\sqrt{2\omega}}\, \hat{L}$ 
 and its leading order approximation for the case $\alpha = \delta=1/2$. 
The exponential decay has been removed.} 
 \label{fig:Lhat-norm-deexp}
 \end{figure}

The construction described in this section can be illustrated by taking the case $\tau=1$, $\alpha=1/2$, $\delta=1/2$ as a concrete example. In this case, the Laplace transform $\tilde{\varphi}(p)=\e^{-\sqrt{p}}$ may be inverted in closed form to give
\begin{equation}
\varphi(t) =\begin{cases} t^{-3/2} \e^{-1/(4t)}/\sqrt{4\pi}  & t>0\\ 0 & t\le 0\end{cases} 
\label{eq:phi}
\end{equation}
See, e.g., formula 17.13.31 in Ref.~\cite{G&R}. 
As $\delta$ has been taken slightly larger than the first maximum of $\varphi$ (which occurs at $\delta=1/6$) the resulting function 
\begin{equation}
H(t) = \varphi(t+1/2)\varphi(1/2-t) = \begin{cases} 
\frac{2}{\pi} (1-4t^2)^{-3/2}\e^{-1/(1-4t^2)}
& |t|<\frac{1}{2} \\ 0 & |t|\ge \frac{1}{2}\end{cases}
\end{equation}
has a broad peak with local minimum at $t=0$ (Fig.~\ref{fig:H}). 
The Fourier transform of $H$, and its leading order asymptotic approximation
\begin{equation}\label{eq:Hhat-specific}
\hat{H}(\omega)\sim \e^{-1/4}\pi^{-1/2}
\e^{-\sqrt{|\omega|/2}}
\cos\left(\frac{|\omega|}{2}-\sqrt{\frac{|\omega|}{2}}\right) \qquad (|\omega|\to\infty)
\end{equation}
are plotted in Fig.~\ref{fig:Hhat}; as the graphs become indistinguishable even at parameters of $|\omega|\sim 20$, they have been plotted again in Fig.~\ref{fig:Hhat-deexp}, over a larger parameter range, but with the leading exponential decay factor divided out. 
The function $L$ defined by
\eqref{eq:Ldef} has integral $0.0658$,\footnote{The numerical method used was to evaluate $101$ samples of $L$ at equally spaced points between $0$ and $1$ inclusive, using numerical integration. The integral of $L$ over $[0,1]$ was evaluated by interpolating a piecewise linear spline through the samples and integrating numerically; the result was doubled to obtain the full integral of $L$. These computations were performed in Maple 18.} and
when one normalizes $L$ so that it has unit integral (and hence $\hat{L}(0)=1$) 
we obtain the function plotted in Fig.~\ref{fig:L-norm}, which has $L(0)=1.4990$. 
Near the switch-on and switch-off regions, i.e., $|t|\sim 1$, we have
\begin{equation}\label{eq:Lt-switching}
L(t)\sim \text{const}\cdot \theta(1-|t|)\sqrt{1-|t|}\e^{-1/(4(1-|t|))}.
\end{equation}
The (normalized) transform is plotted in Fig.~\ref{fig:Lhat-norm} and obeys
\begin{equation}
\hat{L}(\omega) = 2.9324 \,\e^{-\sqrt{2|\omega|}} + O\left(\frac{\e^{-\sqrt{2|\omega|}}}{\sqrt{|\omega|}}\right),
\end{equation}
which is seen to become quite accurate even at moderate values of $|\omega|$. This impression is supplemented by 
Fig.~\ref{fig:Lhat-norm-deexp} in which the same two functions are plotted over a different parameter
range and with the leading exponential decay divided out -- thus the leading order asymptotic contribution in this figure is constant. 

Johnson~\cite{Johnson} has discussed a class of related functions which can be simply defined in coordinate space,
and whose Fourier transforms obtained by a saddle point approximation have a similar asymptotic form to \eqref{eq:Hhat-specific}, but with
an additional power-law decay factor.

\subsection{An electrical model}
 
As an example of how a leading edge of the form~\eqref{eq:Lt-switching} might arise
in controlling a physical device, consider a wire with uniform resistance per unit length $\rho$ and capacitance per unit length $c$ relative to a ground potential. At time $t$ and position $x$ along the wire, let $I(t,x)$ be the current flow in the direction of increasing $x$, and let $V(t,x)$ be the potential relative to ground. Elementary application of the laws of Ohm, Kirchhoff and Volta gives
\begin{equation}
\frac{\partial V}{\partial x} = -I\rho, \qquad \frac{\partial V}{\partial t}= -\frac{1}{c}\frac{\partial I}{\partial x},
\end{equation}
which combine to show that $V$ obeys the diffusion equation
\begin{equation}
\frac{\partial V}{\partial t} = \frac{1}{\rho c} \frac{\partial^2 V}{\partial x^2}.
\end{equation} 
This can be solved for $t,x>0$ with specified boundary condition $V(t,0+)=V_0(t)$, representing the applied control voltage, 
and with $V(0+,x)=0$ for all $x$. Solving by Laplace transform in $t$
and discarding the solution growing in $x$, one finds
\begin{equation}
V(t,x) = \int_0^t G(t-t'; x) V_0(t'), \qquad \tilde{G}(p;x)=  \e^{-x\sqrt{p\rho c}}.
\end{equation}
Note that $\tilde{G}(p;x)=\varphi(p)$, with $\tau=x^2\rho c$. 
In a model where the control voltage is applied sharply at $t=t_*$, 
$V_0(t)=V_*\vartheta(t-t_*)$ and one finds the exact solution
\begin{equation}
V(t,x) = V_*\erfc\left(\sqrt{\frac{x^2\rho c}{4(t-t_*)}}\right) \sim V_* \sqrt{\frac{4(t-t_*)}{\pi \rho c x^2}}  
\e^{-\rho c x^2/(4(t-t_*))}
\end{equation}
for $t\to t_*+$, where $\erfc$ is the complementary error function. 
Comparison with the approximation~\eqref{eq:Lt-switching} shows that
our specific function $L(t)$ for $\alpha=1/2$ gives
a plausible model for the rise of a control voltage at some position
$x$ along the wire. Of course, there are limitations to this model, 
because one cannot model the switch-off in the same way. However,
it indicates that the sampling functions used here are less artificial than
might be thought at first sight. 

\subsection{Relation to positive stable distributions and Fox $H$-Functions}\label{sec:Fox}

We note some further special properties of the functions $\varphi$
used as the basis of our construction. Recall that $\varphi$ is a function on the positive half-line with Laplace transform $\exp(-(p\tau)^\alpha)$. 
One may show that $\varphi$ is a nonnegative function with unit integral over $\RR^+$ (see, e.g., Ref.~\cite{Pollard}) -- in other words, it is a probability
density function. Fixing $\alpha$, the probability density function for the sum of two random variables with densities defined by $\tau_1$ and $\tau_2$ is given by the half-line convolution of their density functions, 
which therefore has Laplace transform $\exp(-(p\tau_3)^\alpha)$ where
$\tau_3=(\tau_1^\alpha+\tau_2^\alpha)^{1/\alpha}$. This is another
distribution in the same family and for this reason, in conjunction with 
their support on the positive half-line, they are called \emph{positive stable distributions}.   

We have already given the explicit form of $\varphi(t)$ in the case $\alpha=1/2$; it is worth noting that the distributions for other values of $\alpha$ can be expressed in terms of Fox $H$-functions. Consider the one parameter family of functions, $g_\alpha(t)$, whose 
Laplace transforms are of the form $\e^{-p^\alpha}$,
\begin{equation}
\int_0^\infty g_\alpha(t)\, \e^{-p t}\, dt = \e^{-p^\alpha}\,.
\end{equation}
A series expansion for $g_\alpha(t)$ was obtained by Humbert~\cite{Humbert} and
proved rigorously by Pollard~\cite{Pollard}: 
\begin{equation}\label{eq:Humbert-Pollard}
g_\alpha(t) = -\frac{1}{\pi}\sum_{k=1}^\infty \frac{(-1)^k}{k! t^{\alpha k+1}}\Gamma(\alpha k+1)\sin \pi k\alpha\,.
\end{equation}
Schneider~\cite{Schneider} has discussed these functions, and shown that
\begin{equation}
g_\alpha(t) = \frac{1}{\alpha\, t^2}\, 
H^{10}_{11}\left( \frac{1}{t} \left| 
\begin{array}{cc}  -1 & 1 \\ {-\alpha^{-1}} & \alpha^{-1} \end{array}  \right.  \right)  \,,
\end{equation}
where $H^{10}_{11}$ is a Fox $H$-function (see, for example, Ref.~\cite{H-fnts}). 
The asymptotic form as $t\rightarrow 0^+$ is 
\begin{equation}
g_\alpha(t) \sim D\, t^{-\mu} \e^{-w\, t^{-\nu}}\,,
\label{eq:f}
\end{equation}
where 
\begin{equation}
\nu = \frac{\alpha}{1-\alpha} \,,
\label{eq:nu}
\end{equation}
\begin{equation}
\mu =  \frac{2-\alpha}{2(1-\alpha)} \,
\label{eq:mu}
\end{equation}
\begin{equation}
w = (1-\alpha)\, \alpha^{\alpha/(1-\alpha)}  \,,
\end{equation}
and
\begin{equation}
D =\frac{\alpha^{1/[2(1-\alpha)]}}{\sqrt{2\pi(1-\alpha)}}   \,.
\end{equation}
Thus for the case $\alpha =1/2$, we have $\nu = 1$, $\mu = 3/2$,   $w=1/4$, and $D=1/\sqrt{4\pi}$ so that
$g_{1/2}(t)$ has the same switch-on behavior as that of $\varphi(t)$ given in Eq.~\eqref{eq:phi}; indeed, $g_{1/2}$ and $\varphi$ coincide.
For more general values of $\alpha$, the switch-on rate is characterized by the parameters
$\nu$ and $\mu$, which in turn are determined by Eqs.~\eqref{eq:nu} and \eqref{eq:mu}. 

At large values of $t$, it is clear from Eq.~\eqref{eq:Humbert-Pollard} that $g_\alpha(t)$ decays slowly -- the distribution has a heavy or fat tail, and indeed, for $0<\alpha<1$,  $g_\alpha$ does not even possess a finite mean. 

Finally, we mention that the functions $g_\alpha$ also appear in
the consideration of fractional derivative extensions of the diffusion equation, formally $\partial^{2\alpha}P/\partial t^{2\alpha}=K\partial^2P/\partial x^2$. Thus the electrical model
of the previous subsection could in principle be generalized to 
a rather broader class of signalling problems in time-fractional diffusion-wave equations (see, e.g.,~\cite{Mainardi}) thus modelling the
switch-on process by an anomalous diffusion processes.

\section{High Moments}
\label{sec:large}

In this section, we will give some general formulas for moments of quadratic operators,
with special attention to the asymptotic forms for high moments. Let $T$ be a normal
ordered quadratic operator which has been averaged in time with a sampling function.
In general, we can expand $T$ as a mode sum
\begin{equation}
T = \sum_{i\, j} (A_{i j}\, a^\dagger_i \,a_j + B_{i j}\, a_i \,a_j + 
B^*_{i j} \, a^\dagger_i \,a^\dagger_j ) \,,
\label{eq:T}
\end{equation}
where $A_{i j}$ and $B_{i j}$ are symmetric matrices, and $a^\dagger_i$ and  $a_i$
are the usual creation and annihilation operators for mode $i$ of a bosonic field. (The expression as a discrete sum is for convenience of presentation.) We define
the moments of $T$ by
\begin{equation}
\mu_n = \langle T^n \rangle \,,
\end{equation}
where the expectation value is in the vacuum state. Thus $\mu_0 =1$, $\mu_1 =0$, and
formulas may be derived for higher moments, such as
\begin{equation}
\mu_2 = 2 \sum_{i\, j} B_{i j}\, B^*_{j i} \, ,
\label{eq:m2}
\end{equation}
\begin{equation}
\mu_3 = 4 \sum_{i\, j\,k}  B_{j i}\, A_{i k}\, B^*_{k j} \,,
\label{eq:m3}
\end{equation}
and
\begin{equation}
\mu_4 = 3\, \mu_2^2 + \sum_{i\, j\,k\, \ell} (16\, B_{j i}\, B_{k \ell} \, B^*_{i k}\,B^*_{j \ell}
+ 4\, B_{j i}\,A_{i k}\, A_{j \ell}\,B^*_{k \ell} + 4\, B_{j i}\,A_{i k}\, A_{k \ell}\,B^*_{j \ell})\,.
\label{eq:m4}
\end{equation}

Consider the example of a massless scalar field, $\varphi$. The mode expansion for
its time derivative at a fixed spatial point can be taken to be
 \begin{equation}
\dot{\varphi} = \sum_{\bf k}\sqrt{ \frac{\omega}{2 V}}\, ( a_{\bf k}\, {\rm e}^{-i \omega t}
  +  a^\dagger_{\bf k}\, {\rm e}^{i \omega t})\,,
\end{equation}
where $\omega = |{\bf k}|$ and $V$ is a quantization volume with periodic boundary
conditions. Let $T$ be the time average with sampling function $f(t)$ of $:\dot{\varphi}^2:$,
so
\begin{equation}
T = \int_{-\infty}^\infty dt\, f(t)\, :\dot{\varphi}^2: \,.
\label{eq:phi2}
\end{equation}
In this case, we find 
\begin{equation}
A_{j \ell} = \frac{\sqrt{\omega_j \omega_\ell}}{V} \, \hat{f}(\omega_j -\omega_\ell )
\label{eq:A}
\end{equation}
and 
\begin{equation}
B_{j \ell} = \frac{\sqrt{\omega_j \omega_\ell}}{2 V} \, \hat{f}(\omega_j +\omega_\ell )\,.
\label{eq:B}
\end{equation}
Here, we have defined the Fourier transform as in Eq.~\eqref{eq:Fourier} and assumed that the sampling function is real and even, $f(t)=f(-t) = f^*(t)$, so its Fourier transform is also real and even,
\begin{equation}
\hat{f}(\omega) = \hat{f}(-\omega) = \hat{f}^*(\omega)\,.
\end{equation}
We will also assume that $f$ is normalized to have unit integral, so
\begin{equation}
\hat{f}(0) =  \int_{-\infty}^\infty dt \, f(t) =1\,.
\end{equation}
Because the summands in the expressions for
$\mu_n$ depend only upon the frequencies of the modes, in the limit of large $V$
we have
\begin{equation}
\sum_{\bf k}  \rightarrow \frac{V}{2 \pi^2} \int_0^\infty d\omega\, \omega^2 \,.
\end{equation}
In this limit, the factors of $V$ cancel, and the moments become integrals of
products of $\hat{f}$ factors.

In the case of a Lorentzian sampling function of width $\tau$,
\begin{equation}
f(t) = f_L(t) = \frac{\tau}{\pi(t^2 +\tau^2)}\,,
\end{equation}
we have
\begin{equation}
\hat{f}(\omega) = \hat{f}_L(\omega) =  \e^{- \tau |\omega|}\,.
\label{eq: LorFT}
\end{equation}
If we use this form, the expressions for the $\mu_n$ given above reproduce the values
given in Table~I in Ref.~\cite{FFR12}.

\subsection{The Dominant Contribution to High Moments}
\label{sec:dom}

As $n$ increases, the number of terms in expressions for $\mu_n$, such as Eq.~\eqref{eq:m4}
grows rapidly. Fortunately, there seems to be one term which gives the dominant contribution.
For $n\ge 3$, this term is 
\begin{equation}
M_n = 4 \sum_{j_1 \cdots j_n} B_{j_1 j_2}\, A_{j_2 j_3}\, A_{j_3 j_4} 
\cdots  A_{j_{n-1} j_n}\,B^*_{j_n j_1}\,,
\label{eq:dom}
\end{equation}
while for $n=2$ the leading factor is $2$. 
For the case $n=4$, this term is the last one in Eq.~\eqref{eq:m4}. We can numerically test
the hypothesis that $\mu_n \sim M_n$ for large $n$ using the case of Lorentzian
sampling of $:\dot{\varphi}^2:$, and the data in Table~I in Ref.~\cite{FFR12}.
 In this case of $n=4$, we find $M_4/\mu_4 \approx 0.843$.
However, this ratio steadily rises toward unity as $n$ increases, becoming $0.966$ at  $n=10$
and $0.993$ at  $n=20$. 

We can understand the origin of the above expression as follows: In every term  of $T^n$ contributing to $\mu_n$, there must always be a
factor of $B_{j_1 j_2}$ on the left and one of $B^*_{j_n j_1}$ on the right. This arises because 
$B_{i j}\, a_i \,a_j$ is the only term in Eq.~\eqref{eq:T} which does not annihilate the vacuum
from the right, and $B^*_{i j}\, a^\dagger_i \,a^\dagger_j $ is the only term which does not
annihilate it from the left. The overall factor of $4$ in Eq.~\eqref{eq:dom} is a combinatorial
factor coming from the two ways to order the ${j_1 j_2}$ indices on the left, and the two ways
to order ${j_n j_1}$ on the right. After the two $B$ factors, there are $n-2$ remaining factors,
which can be combinations of $A$ and $B$ factors. However, the relative minus sign between 
the terms in the argument of $\hat{f}$ in Eq.~\eqref{eq:A} tells us that $A$ tends to fall more
slowly with increasing $\omega_i$ than does $B$. Thus the largest contribution arises when
all of these  $n-2$ factors are those of  $A$, as in Eq.~\eqref{eq:dom}.
In addition to the contribution $M_n$, which can also be written 
as a trace $M_n=4\Tr B A^{n-2} B^*$, there are a number of other 
terms in which the factors in the trace are reordered; these are also
suppressed relative to $M_n$.  In any case, all of the
remaining terms in  $\mu_n$ are nonnegative, so Eq.~\eqref{eq:dom} gives a lower bound on
the moments, $M_n \leq \mu_n$. 

 We may use Eqs.~\eqref{eq:A} and \eqref{eq:B} to write an expression for $M_n$ in
 terms of Fourier transforms of the sampling function. In the limit $V \rightarrow \infty$,
 the result will be of the form
 \begin{equation}
M_n = C_n \int_0^\infty d\omega_1 \cdots d\omega_n (\omega_1 \cdots \omega_n)^p\,
\hat{f}(\omega_1 +\omega_2 ) \hat{f}(\omega_2 - \omega_3 ) \cdots
\hat{f}(\omega_{n-1} - \omega_n ) \hat{f}(\omega_n + \omega_1 ) \,.
\label{eq:Mn} 
\end{equation} 
Here 
\begin{equation}
C_n = \frac{1}{(2 \pi^2)^n}
\label{eq:Cn}
\end{equation}
and $p=3$ for the case of $\dot{\varphi}^2$ given in 
Eq.~\eqref{eq:phi2}, which will be our primary example.  
Equation~\eqref{eq:Mn} can also be obtained as one of the terms 
appearing in the evaluation of $\mu_n$ by Wick's theorem,
as in Sec. III.A and Appendix~B of \cite{FFR12}, where it was identified as the dominant contribution in the case of Lorentzian smearing. 

\subsection{Asymptotic Form of the Moments}
\label{sec:asym}

In this subsection, we develop a procedure for finding asymptotic limits of $M_n$
for large $n$. Consider the case where $n$ is even, so we may write $n = 2m$ and
rewrite the integral in Eq.~\eqref{eq:Mn} as
\begin{equation}
M_{2m} = C_{2m}   \int_0^\infty d\omega_1 d\omega_{m+1} \, (\omega_1 \omega_{m+1})^p\, 
[G_{m-1}(\omega_1, \omega_{m+1})]^2 \,,  
\end{equation}
where $G_{m-1}$ is defined by an $m-1$~dimensional integral,
\begin{equation}
G_{m-1}(\omega_1, \omega_{m+1}) =  \int_0^\infty d\omega_2 \cdots d\omega_m \,
(\omega_2 \cdots \omega_m)^p\,
\hat{f}(\omega_1 +\omega_2 ) \hat{f}(\omega_2 - \omega_3 ) \cdots
\hat{f}(\omega_m - \omega_{m+1} ) \,.
\label{eq:Gdef}
\end{equation}
Note that in general, $G_{m-1}$ will not be symmetric in its two arguments.
We can also write an expression for a general value of $n$ as
\begin{equation}
M_{m+m'} = C_{m+m'}   \int_0^\infty d\omega\, d\Omega\, (\omega \Omega)^p \,
G_{m-1}(\omega, \Omega) \, G_{m'-1}(\omega, \Omega)\,.
\label{eq:M}
\end{equation}
The definition, Eq.~\eqref{eq:Gdef}, leads to a recurrence relation
\begin{equation}
G_\ell(\omega,\Omega) =  \int_0^\infty d\xi \, \xi^p \, \hat{f}(\Omega-\xi)\, 
G_{\ell-1}(\omega,\xi)\,, \quad \ell \geq 1 \,, 
\label{eq:Gn}
\end{equation}
with $G_0(\omega,\Omega) =  \hat{f}(\omega+\xi)$.  
 
Thus, knowledge of $G_1$ allows us to compute all of the $G_\ell$, and hence the
dominant contribution to the moments, $M_n$, for all $n$.

We are especially interested in the case of compactly supported sampling functions, such as
those discussed in Sect.~\ref{sec:compact}. If we adopt units in which $\tau =1$, the 
Fourier transform, $\hat{f}(\omega)$, will fall faster than any power of $\omega$ when
$\omega \gg 1$. We will assume that it falls as an exponential of $\omega^\alpha$, where
$0< \alpha < 1$, specifically that
\begin{equation} 
\hat{f}(\omega) = \gamma\, \e^{-\beta \,|\omega|^\alpha} +O\left(\frac{\e^{-\beta|\omega|^\alpha}}{|\omega|^{1-\alpha}}\right)
\label{eq:f-asy}
\end{equation}
for $\omega\neq 0$. 
We are interested in the asymptotic forms of $G_1$, $G_m$, and finally of $M_n$.
To start, let us proceed heuristically. As $\hat{f}$ becomes more flat at large values of its argument, a reasonable conjecture
is that 
\begin{equation}
G_1(\omega,\Omega) \simeq 2\pi f(0)\, \Omega^p\, \hat{f}(\omega+\Omega) \,.
\label{eq:G1-dom}
\end{equation}
Here $f(0)$, the sampling function evaluated at $t=0$, is expected to be of order unity.
In making this approximation, it is important that our the assumption \eqref{eq:f-asy} entails that $\hat{f}$ does not oscillate at leading order 
(cf.\ the behaviour of the function $\hat{H}$ in Eq.~\eqref{eq:simpleapprox}); if it did,
one might expect a similar approximation, but modified by a factor which would feed through
to later estimates. Proceeding with the conjecture~\eqref{eq:G1-dom}, we may now use the recurrence relation, Eq.~\eqref{eq:Gn}, to find 
a corresponding conjecture for the asymptotic form
for $G_m$. We make the ansatz that 
\begin{equation}
G_{m}^{(p)}(\omega,\Omega) \simeq [2\pi f(0)]^m\, \Omega^{mp}\,  \hat{f}(\omega+\Omega) \,.
\label{eq:Gm-ans} 
\end{equation}
where we have added a superscript to denote the value of $p$. Insertion of this ansatz in
Eq.~\eqref{eq:Gn} leads to
\begin{equation}
G_{m+1}^{(p)}(\omega,\Omega)  \simeq [2\pi f(0)]^m\,  \int_0^\infty d\xi \, \xi^{(m+1)p} \, \hat{f}(\Omega- \xi)\,
\hat{f}(\omega+\xi) =  [2\pi f(0)]^m\,  G_{1}^{((m+1)p)}(\omega,\Omega) \,.
\end{equation}
Now the asymptotic form, Eq.~\eqref{eq:G1-dom}, leads to
\begin{equation}
G_{m+1}^{(p)}(\omega,\Omega) \simeq [2\pi f(0)]^{m+1}\, \Omega^{(m+1)p}\,  \hat{f}(\omega+\Omega) \,.
\end{equation}
which proves our ansatz by induction at this (currently heuristic) level.

In Appendix~\ref{appx:Gm-asy}, these conjectures will be established rigorously. 
To be precise, it will be shown that for each $p\ge 1$ and $m\ge 1$ there exist polynomials $P_m^{(p)}$ 
and $Q^{(p)}_m$ of degrees $mp-1$ and $mp$ respectively, with nonnegative coefficients
independent of $\omega$ and $\Omega$, such that 
\begin{equation}\label{eq:Gm-precise}
\left| 
G_{m}^{(p)}(\omega,\Omega) - [2\pi f(0)]^m\, \Omega^{mp}\,  \hat{f}(\omega+\Omega) 
\right|
\le \left( P_m^{(p)}(\Omega)+\frac{Q_m^{(p)}(\Omega)}{(\omega+\Omega)^{1-\alpha}}\right)
\e^{-\beta(\omega+\Omega)^\alpha}
\end{equation}
for all $\omega,\Omega>0$. This shows that the error terms in
the approximation \eqref{eq:Gm-ans}
are suppressed relative to the dominant term. For instance, if  $\Omega \rightarrow \infty$, with $\omega/\Omega$ fixed, the error term is suppressed by a factor $\Omega^{\alpha-1}$ 
relative to the main term. The statement \eqref{eq:Gm-precise} also holds if $m=0$, 
under the convention that a polynomial of degree $-1$ vanishes identically; in this
case we simply reproduce the assumption \eqref{eq:f-asy} on $\hat{f}$.
Although we do not control the polynomials $P^{(p)}_m$ and
$Q^{(p)}_m$ explicitly (for example, to show that the coefficients are independent of $m$), we do show that there are positive constants $A$, $B$, $c$ and $D$ such that
\begin{equation}\label{eq:rough}
 AB^m\Omega^{mp}\,\e^{-(\omega+\Omega)^{\alpha}} \le 
G^{(p)}_m(\omega,\Omega) \le (cD^m\Omega^{mp}+R^{(p)}_m(\Omega))\e^{-(\omega+\Omega)^{\alpha}} 
\end{equation}
for all $\omega,\Omega>0$, 
where $R^{(p)}_m$ is a polynomial of degree $mp-1$. These
rough bounds provide further evidence that 
\eqref{eq:Gm-ans}  is robust.

Now we seek the asymptotic form for $M_n$ when $n \gg 1$. Use the asymptotic form
for $G_m$, Eq.~\eqref{eq:Gm-ans}, in  Eq.~\eqref{eq:M} to write 
\begin{equation}
M_n \simeq C_n\, [2\pi f(0)]^{n-2}\,  \int_0^\infty d\omega\, d\Omega\, \omega^p \Omega^{(n-1)p} \,
\hat{f}^2(\omega+\Omega) \,, 
\end{equation}
where $n =m +m'$. Next let $u=\omega+\Omega$, and write
\begin{equation}
M_n \simeq C_n\, [2\pi f(0)]^{n-2}\,  \int_0^\infty du\, \hat{f}^2(u) \,   \int_0^u 
d\Omega\, (u-\Omega)^p \, \Omega^{(n-1)p} \,, 
\end{equation}
and then let $x=\Omega/u$ to obtain
\begin{equation}
M_n \simeq C_n\, [2\pi f(0)]^{n-2}\,  \int_0^\infty du\, \hat{f}^2(u) \, u^{np+1}
\,  \int_0^1 dx (1-x)^p\, x^{(n-1)p} \,. 
\end{equation}
The integration on $x$ may be performed to yield 
\begin{equation}
M_n \simeq C_n\, [2\pi f(0)]^{n-2}\,  \frac{p! [(n-1)p]!}{(np+1)!} \;  
 \int_0^\infty du\, \hat{f}^2(u) \, u^{np+1} \,. 
 \label{eq:Ma1}
\end{equation}
If $n \gg 1$, the factor of $u^{np+1}$ will cause the dominant contribution to the 
$u$-integration to come from the region where $u \gg 1$, so we may use the
asymptotic form, Eq.~\eqref{eq:f-asy} to find
\begin{equation}
M_n \simeq C_n\, [2\pi f(0)]^{n-2}\,  \frac{p! [(n-1)p]!}{(np+1)!} \; 
\frac{\gamma^2}{\alpha \,(2 \beta)^{(np+2)/\alpha}}\, \Gamma\left[\frac{(np+2)}{\alpha}\right]\,.
 \label{eq:Ma}
\end{equation}
The most important part of this expression is the final gamma function factor, which leads a rapid
rate of growth of the high moments, $M_n \propto (p n/\alpha)!$.
This result will be used in Sec.~\ref{sec:tail} to infer the tail of the probability distribution. Although we have not performed rigorous
error estimates here, they could be done; however, the
rough bounds \eqref{eq:rough} show that the gamma function factor will be a robust feature of the result. 

The argument given here and in Appendix~\ref{appx:Gm-asy} does not apply to the case of a
Lorentzian sampling function, where $\alpha =1$, as the error term in Eq.~\eqref{eq:Gm-ans} will now be
as large as the dominant contribution. A different argument specifically for the Lorentzian
case is given in Appendix~\ref{appx:Lor}. This argument reproduces the asymptotic form of the 
moments found in  Ref.~\cite{FFR12}.

\subsection{Numerical Tests}
\label{sec:num}

Here we discuss some numerical tests of the asymptotic forms derived in the previous subsection, focussing 
in particular on the form of $G_1(\omega,\Omega)$ for large $\Omega$ with fixed 
$\omega/\Omega$, given in Eq.~\eqref{eq:G1-dom}.  Here we wish to test this result in the case that
$\hat{f}(\omega)$ is exactly given by Eq.~\eqref{eq:f-asy}, with $\beta=\gamma =1$, so
\begin{equation}
\hat{f}(\omega) =  \e^{-|\omega|^\alpha} \,.
\label{eq:alpha}
\end{equation}
Now the exact form for $G_1(\omega,\Omega)$ may be expressed as
\begin{equation}
G_1(\omega,\Omega) = \Omega^{p+1}  \int_0^\infty d\nu \, \nu^p \,
 \exp\left\{-\Omega^\alpha \left[\left(\frac{\omega}{\Omega}+\nu \right)^\alpha +|1-\nu|^\alpha  \right]\right\}\,, 
 \label{eq:G1-2}
\end{equation}
where $\nu = \xi/\Omega$. If we use $f(0) = (2\pi)^{-1} \int_{-\infty}^\infty d\omega\, \hat{f}(\omega) = 
\Gamma(1/\alpha)/(\pi \alpha)$, then the asymptotic form in Eq.~\eqref{eq:G1-dom} may be written
as
\begin{equation}
G_{1A}(\omega,\Omega)  = 
\frac{\Omega^{p}}{\alpha}\, \Gamma(1/\alpha)\, {\rm e}^{-(\omega +\Omega)^\alpha} \,. 
 \label{eq:G1a}
\end{equation} 

\begin{figure}
 \centering
 \includegraphics[scale=0.25]{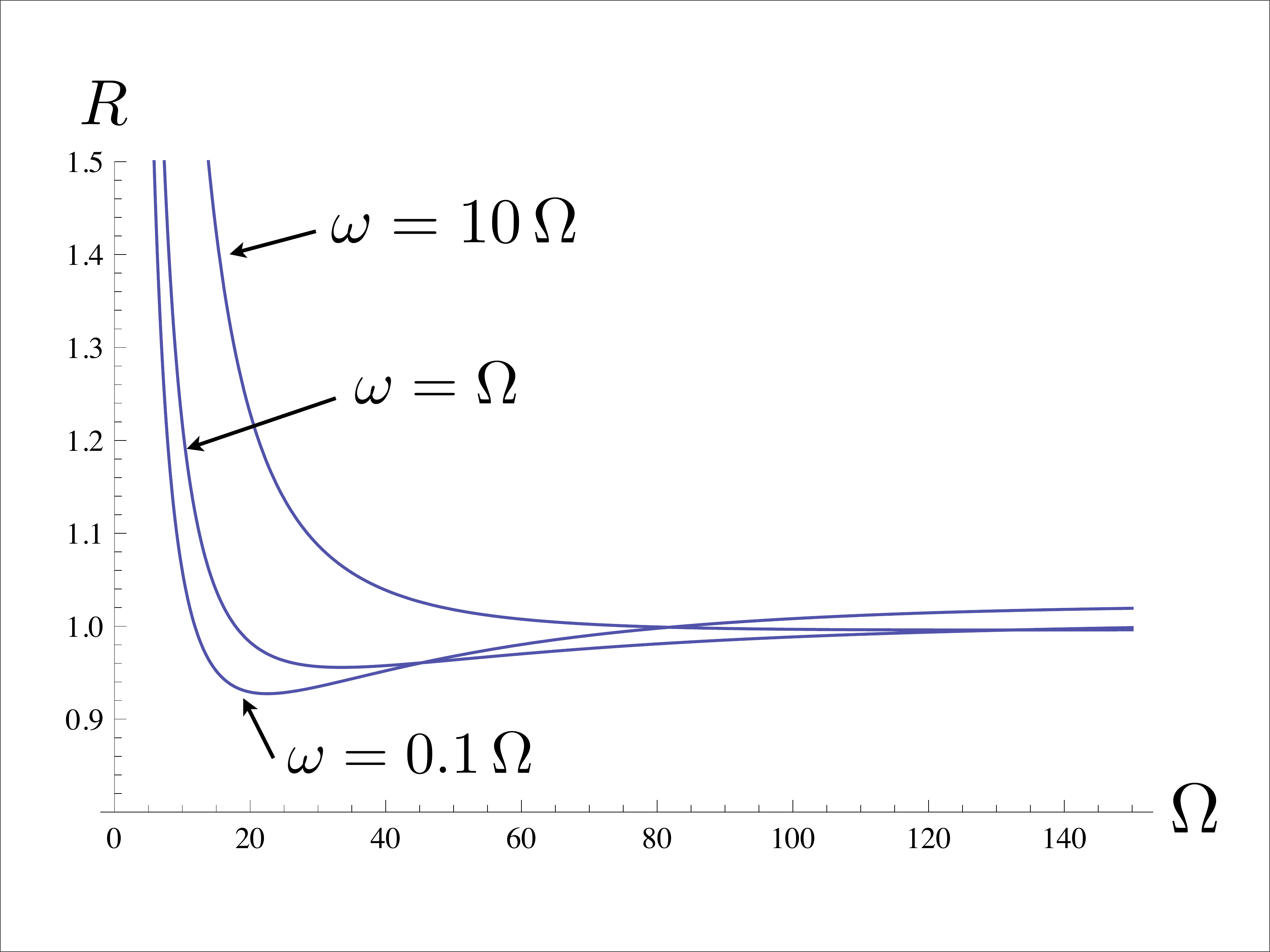}
 \caption{Here $R$, the ratio of the numerical integration for $G_1$, from Eq.~\eqref{eq:G1-2},
 to the asymptotic form, Eq.~\eqref{eq:G1a}, for $p=3$, is plotted for the case $\alpha = 1/2$ as a function
 of $\Omega$ for various values of $\omega/\Omega$.} 
 \label{fig:R-half}
 \end{figure}

\begin{figure}
 \centering
 \includegraphics[scale=0.25]{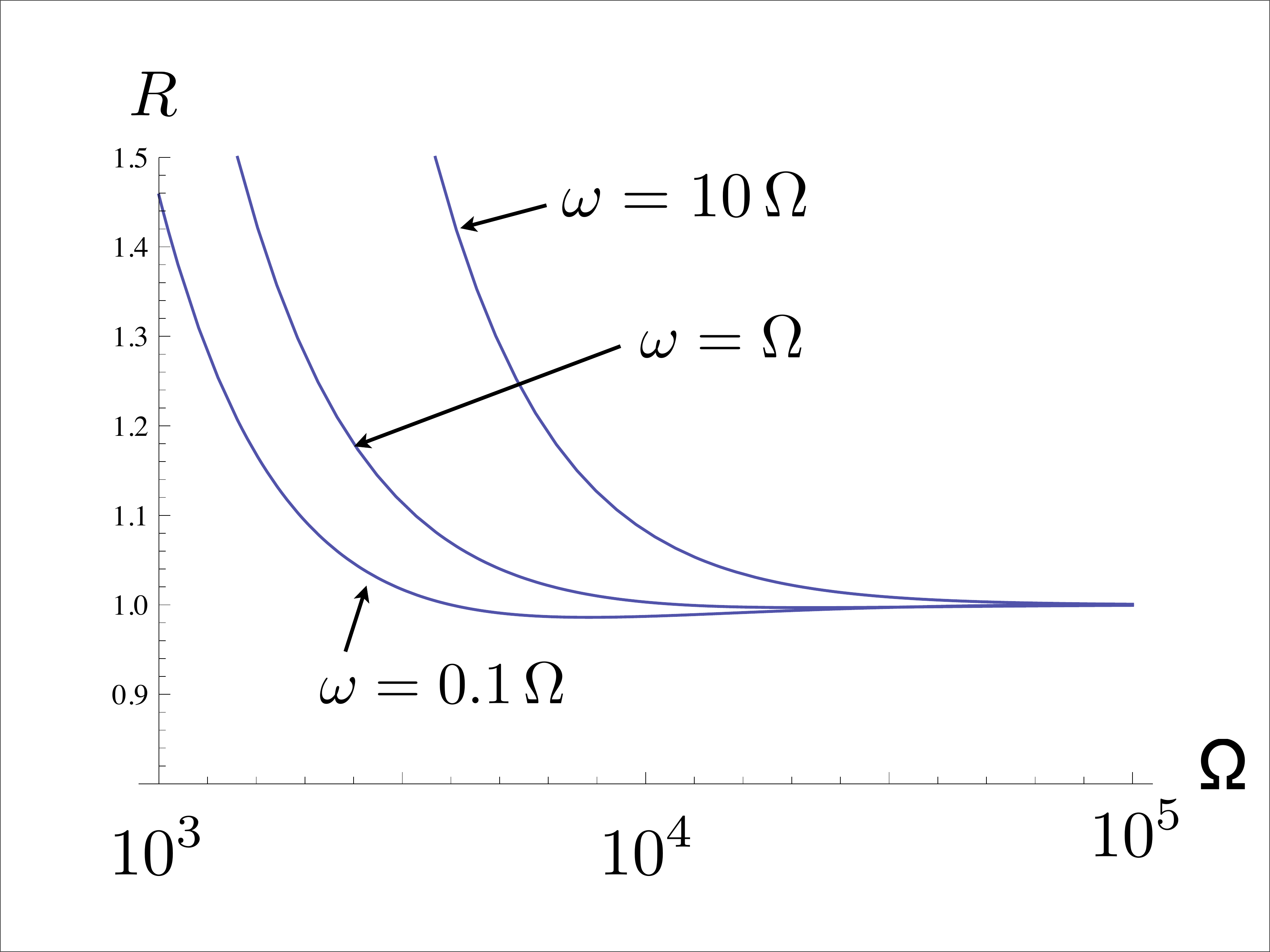}
 \caption{A plot of $R$ for the case   $\alpha = 1/4$. } 
 \label{fig:R-quarter}
 \end{figure}

\begin{figure}
 \centering
 \includegraphics[scale=0.25]{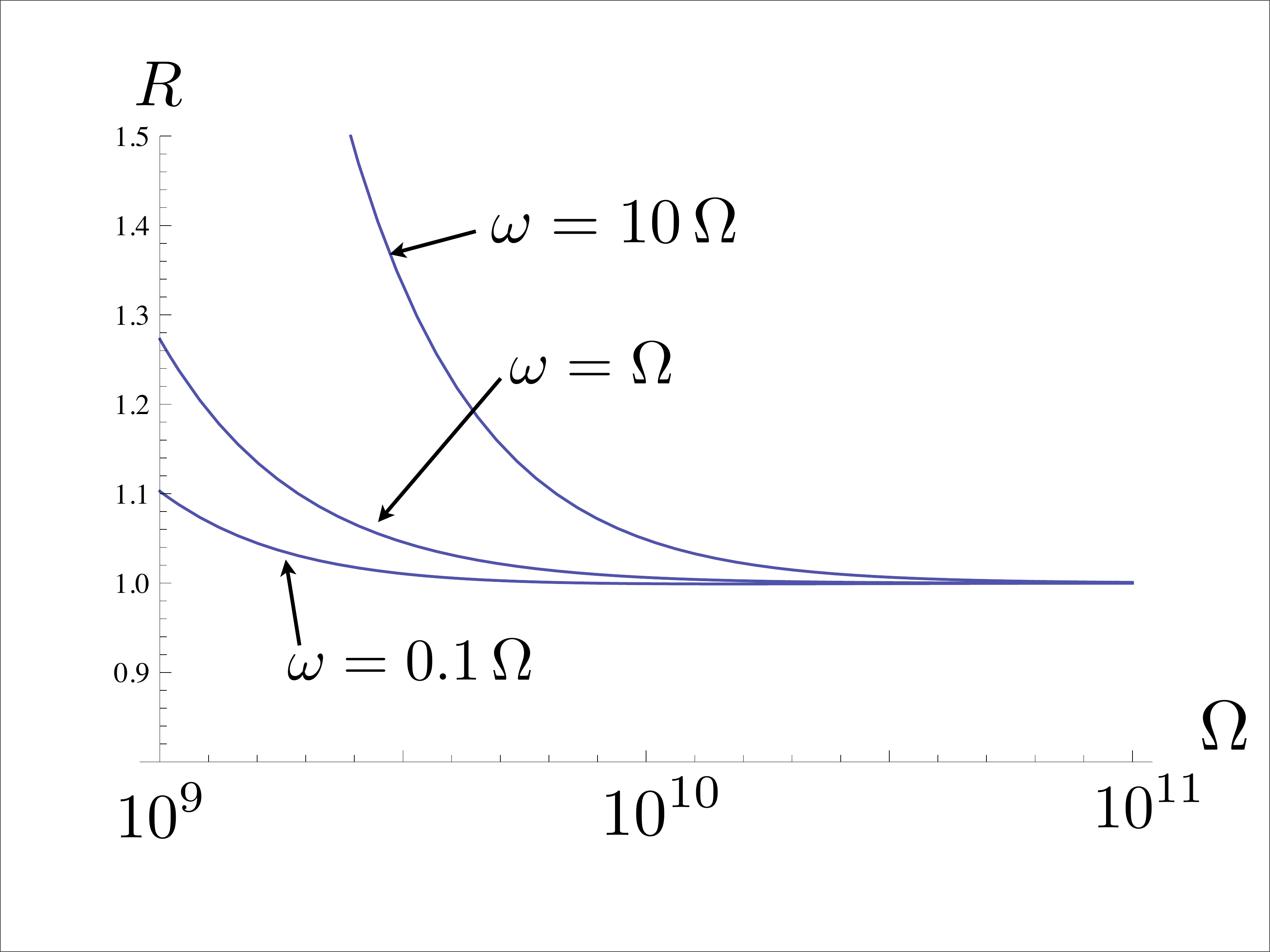}
 \caption{A plot of $R$ for the case   $\alpha = 1/8$. } 
 \label{fig:R-eighth}
 \end{figure}

Let $R$ be the ratio
of the result of numerical integration of Eq.~\eqref{eq:G1-2} to the asymptotic form,  Eq.~\eqref{eq:G1a},
for $p=3$.
In Fig.~\ref{fig:R-half}, this ratio is plotted for the case $\alpha = 1/2$ as a function of $\Omega$ for 
various values of $\omega/\Omega$. We see that $R \rightarrow 1$ for $ \Omega \agt 10^2$, as expected.
(For the case $\omega/\Omega=0.1$, this has not quite occurred on the scale plotted, but does occur
for larger values of $\Omega$.) 
This plot is repeated in Fig.~\ref{fig:R-quarter} for the case $\alpha = 1/4$, where the asymptotic 
limit is reached for $ \Omega \agt 10^5$. The case $\alpha = 1/8$ is illustrated in Fig.~\ref{fig:R-eighth},
where the asymptotic form holds for $ \Omega \agt 10^{10}$. Note that before the asymptotic limit is reached, 
$G_1(\omega,\Omega)$ tends to exceed the asymptotic form, in some cases by large factors. This 
emphasizes that the asymptotic estimates are lower bounds on the exact results. Note that $R\rightarrow 1$
for large $\Omega$ in all of the cases plotted, including when $\alpha < 1/4$. 
The only case plotted where $R < 1$ in part of the range, is in Fig.~\ref{fig:R-half},
when $\omega \alt \Omega$.  However, this effect seems to be small, and does not alter the conclusion that
the asymptotic form,  Eq.~\eqref{eq:G1a}, holds for large $\Omega$.

A separate numerical test can be obtained by numerical evaluation of $M_n$ from Eq.~\eqref{eq:Mn}
using the form of $\hat{f}$ in   Eq.~\eqref{eq:alpha}. Unfortunately, this is only feasible for relatively
small values of $n$. For the case of $p=3$ and $\alpha =1/2$, this evaluation for $\dot{\varphi}^2$ leads to
$M_4 \approx 3.0 \times 10^{12}$ and $ M_6 \approx 1.5 \times 10^{25}$. By comparison, the
asymptotic form, Eq.~\eqref{eq:Ma}, leads to $M_4 \approx 3.0 \times 10^{12}$ and 
$ M_6 \approx 1.0 \times 10^{25}$. These results are in fair agreement, despite the fact that $n=6$
may be too small to expect detailed agreement.

\section{The Tail of the Probability Distribution}
\label{sec:tail}

In this section, we will use the asymptotic form of the moments, Eq.~\eqref{eq:Ma}, to obtain 
information about the probability for large fluctuations. Let the probability distribution, $P(x)$,
be a function of the dimensionless variable $x = T\, \tau^{p+1}$. Our main example is 
$p=3$, so $x = T\, \tau^4$. Recall that in Sec.~\ref{sec:asym}, we computed the asymptotic form
for the moments using units in which $\tau = 1$. We can now restore an arbitrary value for
$\tau$ by using the variable $x$.
We follow the discussion in Sec.~V of Ref.~\cite{FFR12}, and
postulate that the asymptotic form for $P(x)$ for $x \gg 1$ can be written as
\begin{equation}
P(x)  \sim c_0 \,x^b \, {\rm e}^{-a x^c} \,,
\label{eq:pofx}
\end{equation}
for some constants $c_0$, $a$, $b$, and $c$. The $n$-moment for $n \gg 1$ is given by
\begin{eqnarray}
\mu_n &=& \int_{-x_0}^\infty \, x^n \,P(x) \,dx \approx c_0 \, \int_0^\infty \,x^{n+b} \, e^{-a x^c}  \,dx  
\nonumber \\
&=& \frac{c_0}{c} \, a^{-(n+b+1)/c} \, \,  \Gamma[(n+b+1)/c ]  \,.
\label{eq:mun}
\end{eqnarray}
We assume that the asymptotic form of the dominant contribution to $\mu_{n}$, given by
Eq.~\eqref{eq:Ma} will match Eq.~\eqref{eq:mun}, allowing us to infer values for the
constants in Eq.~\eqref{eq:pofx}. Note that the asymptotic form of the moments is growing
too rapidly with increasing $n$ for the moments to define a unique probability distribution
(at least according to the Hamburger or Stieltjes criteria~\cite{Simon}). 
However, the difference between two probability distributions with the same moments must
be an oscillatory function, all of whose moments vanish. We are less interested in the detailed
functional form of $P(x)$ than in the its average rate of decay for large $x$, and we expect this
average rate to be of the form of Eq.~\eqref{eq:pofx}. 
This reasoning can be supported by rigorous arguments -- see 
of Sec.~VI of Ref.~\cite{FFR12}.

First, note that
\begin{equation}
\Gamma\left[\frac{np+2}{\alpha}\right] = (n p/\alpha + 2/\alpha -1)!\,,
\end{equation}
and that 
\begin{eqnarray}
\frac{ [(n-1)p]!}{(np+1)!}&=& \frac{\alpha^{-(p+1)}}{(np/\alpha +1/\alpha)(np/\alpha) \cdots
(np/\alpha -{\frac{p-1}{\alpha}})} \nonumber \\ &\sim&
\frac{\alpha^{-(p+1)}}{(np/\alpha +2/\alpha-1) \cdots
(np/\alpha +2/\alpha-p-1)}\,\left[1 + O\left(\frac{1}{n}\right)\right] \,.
\label{eq:factorials}
\end{eqnarray}
In the last step, we use the fact that $n$ is large, but $p$ is fixed, so
\begin{equation}
\frac{1}{n p +A} \sim \frac{1}{n p }\,\left[1 + O\left(\frac{1}{n}\right)\right] \,,
\end{equation}
for any constant $A$. We assume that $C_n$ is of the form 
\begin{equation}
C_n = B_0 \, B^n
\end{equation}
for constants $B_0$ and $B$. This form holds for the case of $\dot{\varphi}^2$, given in Eq.~\eqref{eq:Cn}.

Next we drop the $O(1/n)$ terms in Eq.~\eqref{eq:factorials} and rewrite $M_n$ in
Eq.~\eqref{eq:Ma} as
\begin{equation}
M_n \sim B_0 \,  p!\, \gamma^2\, \alpha^{-(p+2)} \, B^n\, [2\pi f(0)]^{n-2}\, (2 \beta)^{-(np+2)/\alpha}\,
 \Gamma(np/\alpha +2/\alpha-p-1) \,.
\label{eq:Mn2}
\end{equation}
Now we can compare Eqs.~\eqref{eq:mun} and \eqref{eq:Mn2} to
obtain expressions for $b$ and $c$ in terms of $p$ and $\alpha$ alone:
\begin{equation}
c = \frac{\alpha}{p}
\label{eq:c}
\end{equation}
and
\begin{equation}
b = c\, \left(\frac{2}{\alpha} -p-1\right) -1 
{\color{blue}= \frac{2-\alpha}{p} -(\alpha+1)}\,.
\end{equation}
We also obtain expressions for $a$ and $c_0$ as
\begin{equation}
a = 2\beta\,  [2\pi f(0) B]^{-\alpha/p}\,,
\end{equation}
and
\begin{equation}
c_0 = c\, a^{(b+1)/c}\, B_0\, p!\, \gamma^2 \, \alpha^{-(p+2)} \, (2 \beta)^{-(2/\alpha)}\,
{[2\pi f(0)]^{-2}}\,.
\end{equation}
We are especially interested in operators, such as stress tensors, for which $p=3$, so
$c = \alpha/3$ and $b=-(4\alpha +1)/3$.
Although Eq.~\eqref{eq:Ma} does not strictly hold for the Lorentizian
sampling function, $\alpha =1$, we do reproduce the $c = 1/3$ result of Ref.~\cite{FFR12}.
Compactly supported sampling functions with $\alpha < 1$ lead to an even slower
rate of decrease of the tail, such as c =1/6 for $\alpha = 1/2$. Recall that the
asymptotic form for the high moments which we have used, Eq.~\eqref{eq:Ma},
is expected to be a lower bound on the exact moments. This means that the form of the 
asymptotic probability distribution, Eq.~\eqref{eq:pofx} is a lower bound on the exact
$P(x)$, which could potentially fall even more slowly for large $x$.

\section{Summary and Discussion}
\label{sec:sum}

In this paper, we have studied the probability distributions for quantum stress tensor operators
averaged in time with functions of compact support, which describe measurements in a finite
time interval. Our main interest is in the asymptotic form of the  distribution, which give the
probability for large fluctuations in the vacuum state.  This asymptotic form is determined by
the rate of growth of the high moments of the distribution. This rate of growth was inferred by first
identifying a dominant contribution to the high moments, Eq.~\eqref{eq:dom}, which is determined by 
the Fourier transform of the sampling function. We next used asymptotic analysis to find the
the rate of growth of the dominant contribution, which is given in Eq.~\eqref{eq:Mn2}.  This rate
is typically too fast for the moments to uniquely determine the probability distribution according to
the Hamburger or Stieltjes criteria. However,
asymptotic form for the moments is adequate to give estimates for the probability for large fluctuations,
which was done in Sec.~\ref{sec:tail}. Note that the dominant contribution, $M_n$, is a lower bound
on the exact moments, $\mu_n$, so our estimate of the asymptotic probability distribution is a lower bound
on the exact result and hence on the probability of large fluctuations.

Compactly supported sampling functions have Fourier transforms which fall more slowly than an exponential,
resulting in more rapid growth of the moments compared to the case of Lorentzian averaging, and a
more slowly falling tail of the  probability distribution. For example, the sampling functions with $\alpha = 1/2$,
such as those discussed in Sec.~\ref{sec:examples}, result in moments growing as $(6 n)!$ and  an 
asymptotic probability distribution for large $x$ of the form of
 $x^{-1}\, {\rm e}^{-x^{1/6}}$, where $x$ is the dimensionless measure of the sampled stress tensor. This
slow rate of decrease leads to the possibility of significant physical effects from large fluctuations.
Furthermore, the rate of decrease is very sensitive to the value of $\alpha$, with smaller $\alpha$
leading to an even greater probability of large fluctuations. The value of $\alpha$ is linked to the rate
of switch-on and switch-off of the sampling functions, as described by Eqs.~\eqref{eq:f}, \eqref{eq:nu}, 
and \eqref{eq:mu}.

There are several possible physical applications of these results,  including the effects of quantum
radiation pressure fluctuations on barrier penetration by charged particles or upon lightcone fluctuation effects 
in a nonlinear material. These are quadratic operator fluctuation effects, analogous to the linear electric
field fluctuations studied in Refs.~\cite{BDFS14,HF15}. In both cases, the physical system may define a choice 
of the sampling function  $f_\tau(t)$. For example, the effects of the fluctuations of the squared electric field in a nonlinear
material are determined by an integral along the path of a probe pulse, as described in Ref.~\cite{BDFS14}.
This integral is in turn depends upon the density profile of the slab.   This illustrates the general principle that
physical processes which begin and end at finite times should be described by sampling functions with compact
support, and the specific form of these functions is ultimately determined by the details of the physical process
in question. Because all of the probability distributions treated here fall more slowly than exponentially, even
at finite temperature sufficiently large fluctuations of, for example, energy density are more likely to be vacuum
fluctuations than thermal fluctuations. This and other physical consequences of the results described in this
paper need to be more carefully explored.

\begin{acknowledgments}
We would like to thank Tom Roman for useful discussions in the early stages of this work, and
Haiyun Huang for helpful comments on the manuscript. CJF thanks
Jorma Louko and Benito Ju\'{a}rez Aubry for bringing Ref.~\cite{Ingham} to his attention, and Simon Eveson for suggesting the link with positive stable distributions. 
\
This work was supported in part  by the National Science Foundation under Grant  PHY-1205764.
\end{acknowledgments}

\appendix

\section{Integral estimates}\label{appx:int-ests}

The aim of this appendix is to prove Theorem~\ref{thm:I-ests}. This is accomplished
as a special case of the following result.
\begin{lemma} \label{lem:int-ests}
Suppose that positive functions $\Psi, \Psi_1$ and $\Upsilon$ are defined on $(0,\infty)$, with $\Psi_1$ monotone decreasing, and obeying
\begin{equation}\label{eq:Upsilon}
\int_0^\infty d\omega (1+\omega)\Upsilon(\omega)<\infty  
\end{equation}
and 
\begin{equation}\label{eq:hyps}
\Psi(\omega)\int_{\lambda\omega}^\infty d\omega' \Upsilon(\omega') = O(\Psi_1(\omega)),
\qquad
\int_0^{\lambda\omega} d\omega'\,\omega'\Psi_1(\omega-\omega')\Upsilon(\omega')=O(\Psi_1(\omega))
\end{equation}
for $\omega>0$ and fixed $\lambda\in [0,1]$. Let $\psi:(0,\infty)\to\CC$ be differentiable and $\upsilon:\RR\to\CC$ be continuous, with $\upsilon(\omega)=O(\Upsilon(|\omega|))$ for all $\omega\in\RR$ and suppose there is $\omega_0\ge 0$ so that
$\psi(\omega)=O(\Psi(\omega))$, $\psi'(\omega)=O(\Psi_1(\omega))$ for $\omega>\omega_0$. Then
\begin{equation}\label{eq:conv_est}
\int_{-\lambda\omega}^\infty d\omega'\, \psi(\omega+\omega')\upsilon(\omega')=
\psi(\omega)\int_{-\infty}^\infty d\omega'\,\upsilon(\omega')
+O(\Psi_1(\omega))
\end{equation}
for $\omega>\omega_0/(1-\lambda)$ if $\lambda<1$ or $\omega>0$ if $\omega_0=0$. 
If, additionally, $\psi'$ is differentiable with $\psi''(\omega)=O(\Psi_2(\omega))$ where $\Psi_2$ is a positive, monotone decreasing function
on $(0,\infty)$ so that properties \eqref{eq:hyps} also hold with $\Psi$
and $\Psi_1$ replaced by $\Psi_1$ and $\Psi_2$ respectively, then
\begin{equation}\label{eq:conv_deriv_est}
\frac{d}{d\omega}\int_{-\lambda\omega}^\infty d\omega'\, \psi(\omega+\omega')\upsilon(\omega')=
\psi'(\omega)\int_{-\infty}^\infty d\omega'\,\upsilon(\omega')
-\lambda\psi((1-\lambda)\omega) \upsilon(-\lambda\omega)
+O(\Psi_2(\omega)) 
\end{equation} 
holds for $\omega>\omega_0/(1-\lambda)$ if $\lambda<1$ or $\omega>0$ if $\omega_0=0$.
\end{lemma}
{\noindent\emph{Proof:}} The integrals in Eq.~\eqref{eq:conv_est} are
finite because $\upsilon$ is necessarily absolutely integrable and $\psi$ is bounded.  A straightforward rearrangement of terms gives
\begin{equation}
\int_{-\lambda\omega}^\infty d\omega'\, \psi(\omega+\omega')\upsilon(\omega') -
\psi(\omega)\int_{-\infty}^\infty d\omega'\,\upsilon(\omega')
= -\psi(\omega)\int_{-\infty}^{-\lambda\omega}d\omega' \upsilon(\omega') +
\int_{-\lambda\omega}^\infty d\omega' \left(\psi(\omega+\omega')-\psi(\omega)\right)\upsilon(\omega')
\end{equation}
and the first term on the right-hand side is $O(\Psi_1(|\omega|))$
by the hypotheses. Next, the mean value theorem gives $\psi(\omega+\omega') - \psi(\omega) = \omega' \psi'(\zeta)$ for some
$\zeta$ lying between $\omega$ and $\omega+\omega'$, giving
\begin{equation}
|\psi(\omega+\omega') - \psi(\omega) | 
\le  |\omega'|\, \Psi_1(\min\{\omega,\omega+\omega'\}) 
\end{equation}
on recalling that $\Psi_1$ bounds $\psi'$ and is monotone decreasing.
Splitting the integration range at $\omega=0$,  this gives
\begin{equation}
\left|\int_{-\lambda\omega}^\infty d\omega' \left(\psi(\omega+\omega')-\psi(\omega)\right)\upsilon(\omega')\right|
\le
\Psi_1(\omega) \int_{0}^\infty d\omega'\, \omega'\Upsilon(\omega') + 
\int_0^{\lambda\omega} d\omega'\,\omega' \Psi_1(\omega-\omega')\Upsilon(\omega')
\end{equation}
where we have also changed $\omega'\to-\omega'$ in the last integral.
As both terms on the right-hand side are $O(\Psi_1(|\omega|))$
the first part of the result is proved. 
For the second part, we first note that
\begin{equation}\label{eq:conv_deriv_est2}
\frac{d}{d\omega}\int_{-\lambda\omega}^\infty d\omega'\, \psi(\omega+\omega')\upsilon(\omega')= 
-\lambda\psi((1-\lambda)\omega) \upsilon(-\lambda\omega)
+ \int_{-\lambda\omega}^\infty d\omega'\, \psi'(\omega+\omega')\upsilon(\omega')
\end{equation}
and apply the first part of the lemma to the second term. 
$\square$

The proof of Theorem~\ref{thm:I-ests} now amounts to verifying that 
the hypotheses of Lemma~\ref{lem:int-ests} are satisfied for the functions 
\begin{equation}
\Psi(\omega)=\Upsilon(\omega) = \exp(-\beta\omega^\alpha),\qquad
\Psi_1(\omega) = \frac{\exp(-\beta\omega^\alpha)}{\omega^{1-\alpha}}, 
\qquad
\Psi_2(\omega) = \frac{\exp(-\beta\omega^\alpha)}{\omega^{2-\alpha}}
\end{equation}
where $\beta>0$, $0<\alpha<1$, in the case $\lambda=1/2$. 
Condition \eqref{eq:Upsilon} is obvious, 
while we may also compute
\begin{equation}\label{eq:tail}
\int_{\Omega}^\infty d\omega' \e^{-\beta(\omega')^\alpha}
= \Omega\int_{0}^\infty du\, \e^{-\beta\Omega^\alpha(1+u)^{\alpha}}
\sim \frac{\Omega^{1-\alpha}\e^{-\beta\Omega^\alpha}}{\alpha\beta}
\qquad (\Omega\to\infty)
\end{equation} 
using Laplace's method (see Lemma~\ref{lem:Laplace} below). 
As the right-hand side is bounded for all $\Omega>0$ and
decays rapidly as $\Omega\to\infty$, it follows
that, for each $\gamma>0$, the function $\omega^{\gamma} \int_{\omega/2}^\infty d\omega' \exp(-\beta(\omega')^\alpha)$ is bounded for all $\omega>0$ (with a bound depending on $\gamma$). 
Hence  the first condition in \eqref{eq:hyps} holds for $\Psi$ and $\Psi_1$
and also for $\Psi_1$ and $\Psi_2$. The second condition in  \eqref{eq:hyps} is verified by
noting that
\begin{align}
\int_0^{\omega/2}d\omega'\,\omega' \frac{\e^{-\beta[(\omega-\omega')^\alpha+\omega'{}^\alpha]}}{(\omega-\omega')^\eta} 
&= \frac{\omega^{2-\eta}}{\alpha}\int_0^{2^{-\alpha}} 
dv\,\frac{v^{2/\alpha-1}}{(1-v^{1/\alpha})^\eta}\e^{-\beta\omega^\alpha(v+(1-v^{1/\alpha})^\alpha)}
\end{align}
where we have made the change of variables $v=(\omega'/\omega)^\alpha$. The integral may be estimated by Laplace's method
in the form of Lemma~\ref{lem:Laplace} below; 
in particular, the integral is $O(\omega^{-\eta}\e^{-\beta\omega^\alpha})$ for $\omega>0$.
This calculation shows that the second condition in \eqref{eq:hyps} is verified for both
$\Psi_1$ and $\Psi_2$ for $0<\alpha<1$. Note, however, that this condition would fail 
in the case $\alpha=1$.  Eq.~\eqref{eq:I-omega} in Theorem~\ref{thm:I-ests} is now
immediate, while Eq.~\eqref{eq:Iprime-omega} follows on noting that 
\begin{equation}
\Psi(\omega/2)^2 = \e^{-2\beta(\omega/2)^\alpha} = O(\omega^{\alpha-1}\e^{-\beta\omega^\alpha})
\end{equation}
as $0<\alpha<1$.

It remains to state Laplace's method in the form we use it.
\begin{lemma} \label{lem:Laplace}
Let $V,\gamma>0$ be constant and suppose $g:[0,V]\to\RR$ is continuous with $g(0)\neq 0$ and $h:[0,V]\to\RR$ is continuous, attains its maximum only at $v=0$, and has a strictly negative one-sided derivative $h'(0)$. Then the integral
\begin{equation} 
I(t)= \int_0^V dv\, v^{\gamma-1}g(v) \e^{th(v)}
\end{equation}
obeys  
\begin{equation}\label{eq:Laplace}
I(t) \sim \frac{\Gamma(\gamma)g(0)\e^{th(0)}}{ (t|h'(0)|)^{\gamma}}\qquad
(t\to\infty) 
\end{equation}
and $I(t)=O(t^{-\gamma}\e^{th(0)})$ on $(0,\infty)$.
The asymptotic equivalence~\eqref{eq:Laplace} also holds 
if $[0,V]$ is replaced by $[0,\infty)$, under the
additional assumptions that $h(v)\to -\infty$ as $v\to\infty$ and
that $I(t)$ converges for $t=t_0$; in this case we have
$I(t)=O(t^{-\gamma}\e^{th(0)})$ on $(t_0,\infty)$.
\end{lemma}
\emph{Proof} The asymptotic equivalence holds by a straightforward adaptation of 
the argument in \S4.3 of \cite{deBruijn} (which deals
with the $\gamma=1$ case). As $I(t)$ is evidently bounded on any compact set, we may 
deduce that $I(t) = O(t^{-\gamma}\e^{th(0)})$ on $(0,\infty)$ or $(t_0,\infty)$ as appropriate. $\square$

\section{Asymptotic behavior of $G_m$}\label{appx:Gm-asy}

The aim of this appendix is to prove statement \eqref{eq:Gm-precise} on the behavior of $G_m(\omega,\Omega)$. Throughout, we will assume
\eqref{eq:f-asy}, that  $\hat{f}(\omega) = \gamma\, \e^{-\beta \,|\omega|^\alpha} +O\left(\e^{-\beta|\omega|^\alpha}/|\omega|^{1-\alpha}\right)$, where $0<\alpha<1$ and $\beta,\gamma>0$. 
It will also be useful to note that the magnitude of $\hat{f}(\xi)$  satisfies
\begin{equation}
|\hat{f}(\omega)| \leq c \, {\rm e}^{-\beta \,|\omega|^\alpha} \,.
\label{eq:f-bound}
\end{equation}
for some constant $c $. This follows because $\hat{f}$ is bounded and obeys Eq.~\eqref{eq:f-asy},  although in general $c \not= \gamma$.

\subsection{Preliminary estimates}\label{appx:prelim}

Define a family of integrals
\begin{equation}
I^r_\nu(\omega,\Omega) = 
 \int_0^\infty d\xi\frac{(\Omega+\xi)^r}{(\omega+\Omega+\xi)^\nu}
\e^{-\beta\left(\xi^\alpha+(\omega+\Omega+\xi)^\alpha\right)}\,,
\end{equation}
where $\omega,\Omega>0$, $0\le \nu<1$ and $r\ge 0$ is an integer. Noting that the factor $(\omega+\Omega+\xi)^{-\nu}
\e^{-\beta (\omega+\Omega+\xi)^\alpha}$ is decreasing
on $[0,\infty)$, we obtain the obvious estimate 
\begin{equation}
I^r_\nu(\omega,\Omega)\le \frac{\e^{-\beta (\omega+\Omega)^\alpha}}{(\omega+\Omega)^{\nu}}
J^{0,r}(\Omega)\,,
\end{equation}
where we define, for $q\ge 0$, 
\begin{equation}
J^{q,r}(\Omega) = \int_0^\infty d\xi \, \xi^q (\Omega+\xi)^r 
\e^{-\beta \xi^\alpha},
\end{equation}
which is evidently a polynomial of degree $r$ in $\Omega$ with
nonnegative coefficients (independent of $\omega$) that can 
be evaluated in terms of Euler $\Gamma$-functions: for instance, 
one easily sees that
\begin{equation}
J^{(q,r)}(\Omega) = \frac{\Gamma((q+1)/\alpha)}{\alpha\beta^{(q+1)/\alpha}}\Omega^r + \text{lower order terms}.
\end{equation}
In particular, the leading coefficient is independent of $r$.

Next, let us make the elementary observation that the inequality
\begin{equation}
x^\alpha + y^\alpha \ge (x+y)^\alpha + (1-\alpha)\min\{x,y\}^\alpha
\end{equation}
holds for all $x,y>0$, $0<\alpha<1$. This is seen by the calculation (assuming without loss of generality that $x\le y$)
\begin{equation}
x^\alpha + y^\alpha - (x+y)^\alpha = x^\alpha- \alpha \int_{y}^{x+y} dz\, z^{\alpha-1} \ge x^\alpha - \alpha x y^{\alpha-1}
\ge (1-\alpha)x^\alpha\,,
\end{equation}
where the integral was estimated by mutiplying the measure of the integration range by the maximum value of the integrand; note that
$x^\alpha\le y^\alpha$ but $y^{\alpha-1}\le x^{\alpha-1}$. 

Now consider 
\begin{equation}
K^{q,r}_\nu(\omega,\Omega) =
 \int_0^\Omega d\xi\frac{\xi^q(\Omega-\xi)^r}{(\omega+\Omega-\xi)^\nu}
\e^{-\beta\left(\xi^\alpha+(\omega+\Omega-\xi)^\alpha\right)}\,,
\end{equation}
where $\omega,\Omega>0$, $0\le \nu<1$, $q\ge 0$ and $r\ge \nu$ is an integer. Evidently
\begin{align}
K^{q,r}_\nu(\omega,\Omega)
&\le \frac{\Omega^r}{(\omega+\Omega)^\nu}
\int_{0}^\Omega d\xi \,\xi^q
\e^{-\beta\left(\xi^\alpha+(\omega+\Omega-\xi)^\alpha\right)}
\\
&\le \frac{\Omega^r\e^{-\beta(\omega+\Omega)^\alpha}}{(\omega+\Omega)^\nu}
\int_{0}^\Omega d\xi \,\xi^q
\e^{-\beta(1-\alpha)\min\{\xi,\omega+\Omega-\xi\}^\alpha}\,,
\label{eq:Iqrnu}
\end{align}
where we have used the fact that $(\Omega-\xi)^r/(\omega+\Omega-\xi)^\nu$ is
decreasing on $[0,\Omega]$ because $0\le \nu \le r$.
The last integral can be rewritten
\begin{equation}
\int_{0}^{\min\{\Omega,\frac{1}{2}(\omega+\Omega)\}} d\xi \,
\xi^q
\e^{-\beta(1-\alpha)\xi^\alpha}
+\int_{\min\{\Omega,\frac{1}{2}(\omega+\Omega)\}}^\Omega
d\xi \,\xi^q\e^{-\beta(1-\alpha)(\Omega+\omega-\xi)^\alpha}\,,
\end{equation}
the first term of which can be bounded by a constant independent of $\omega$ (replacing the upper limit by $\infty$). In the second term, we may
estimate the exponential from above by $\e^{-\beta(1-\alpha)(\Omega/2)^\alpha}$, whereupon the integral is bounded above
by $\Omega^{q+1}\e^{-\beta(1-\alpha)(\Omega/2)^\alpha}$ and hence
by a constant independent of $\omega$. Thus the integral in
\eqref{eq:Iqrnu} is bounded by a constant
and we have shown that 
\begin{equation}
K^{q,r}_\nu(\omega,\Omega) =O\left(\frac{\Omega^r\e^{-\beta(\omega+\Omega)^\alpha}}{(\omega+\Omega)^\nu}\right)
\end{equation}
for $\omega,\Omega>0$. Note that the controlling constant in this
estimate (i.e., the supremum over $\Omega>0$ of the integral in formula \eqref{eq:Iqrnu}) is independent of $r$. 

A further family that will appear below is defined by
\begin{equation}
L^r_\nu(\omega,\Omega) = 
 \int_0^\infty d\xi\frac{\xi^r}{(\omega+\xi)^\nu}
\e^{-\beta\left(|\Omega-\xi|^\alpha+|\omega+\xi|^\alpha\right)}\,,
\end{equation}
where, as before, $\omega,\Omega>0$, $0\le \nu<1$ and $r\ge \nu$ is an integer. By changing variable $\xi\mapsto \xi-\Omega$ and then splitting the integration region into $[0,\infty)$ and $[-\Omega,0]$,
we find (changing variables again $\xi\mapsto -\xi$ in the second of these) 
\begin{equation}
L^r_\nu(\omega,\Omega) = I^r_\nu(\omega,\Omega) + K^{0,r}_\nu(\omega,\Omega)
\end{equation}
and hence deduce that 
\begin{equation}\label{eq:Lrnu}
L^r_\nu(\omega,\Omega) \le \frac{P(\Omega)\e^{-\beta(\omega+\Omega)^\alpha}}{(\omega+\Omega)^\nu}\,,
\end{equation}
where $P$ is a polynomial of degree $r$ with nonnegative coefficients
independent of $\omega$ and $\Omega$, and whose leading order
coefficient is independent of $r$.

\subsection{Rough bounds on $G_m$}

The iterative scheme for computing the functions $G_m$ may be written in the form
\begin{equation}
G^{(p)}_{m+1}=\Xi^{(p)}G^{(p)}_m, \qquad
G^{(p)}_0(\omega,\Omega) = \hat{f}(\omega+\Omega)\,,
\end{equation}
where we have inserted a superscript to track the value of $p\ge 1$ and the integral operator $\Xi^{(p)}$ is defined by
\begin{equation}
(\Xi^{(p)} G)(\omega,\Omega)
=\int_0^\infty d\xi\, \xi^p \hat{f}(\Omega-\xi) G(\omega,\xi)\,.
\end{equation} 
Note that, due to the assumption \eqref{eq:f-bound}, we have an estimate
\begin{equation}\label{eq:Xi-ineq}
|(\Xi^{(p)} G)(\omega,\Omega)|
\le c \int_0^\infty d\xi\, \xi^p \e^{-\beta|\Omega-\xi|^\alpha} |G(\omega,\xi)|\,.
\end{equation} 

As a preparation for the main estimate, we first show that  
\begin{equation} \label{eq:Gm-rough-upper}
|G^{(p)}_m(\omega,\Omega)| \le Q^{(p)}_m(\Omega) \e^{-\beta(\omega+\Omega)^\alpha}  
\end{equation}
for $\omega,\Omega>0$,  where $Q^{(p)}_m$ is a polynomial
of degree $mp$ with nonnegative coefficients independent of $\omega$ and $\Omega$.  The statement is true by definition for $m=0$, so suppose it holds for some $m\ge 0$. Using the inductive hypothesis and 
also the estimate \eqref{eq:Xi-ineq} gives
\begin{equation}\label{eq:Gpm-ineq}
|G^{(p)}_{m+1}(\omega,\Omega)|\le c 
\int_0^\infty d\xi\, \xi^pQ^{(p)}_m(\xi)    
\e^{-\beta(|\Omega-\xi|^{\alpha}+|\omega+\xi|^\alpha)}\,,
\end{equation}
which can be expanded as a linear combination of integrals of
the form $L^{r}_0(\omega,\Omega)$ with $p\le r\le (m+1)p$, 
in which the coefficients are positive and independent of $\omega,\Omega$. Using  \eqref{eq:Lrnu},  we deduce 
that $|G^{(p)}_{m+1}(\omega,\Omega)| \le Q^{(p)}_{m+1}(\Omega) \e^{-\beta|\omega|^\alpha}$ where $Q^{(p)}_{m+1}$ is a polynomial
of degree $(m+1)p$ with nonnegative coefficients independent of $\omega$ and $\Omega$. Thus the bound \eqref{eq:Gm-rough-upper} holds for all integers $m\ge 0$ by induction.

In principle, the polynomials $Q^{(p)}_m$ could be computed explicitly, but this would be a matter of some tedium. The leading order coefficient can seen to grow no faster than geometrically with $m$, because
the leading order contribution to the right-hand side of inequality \eqref{eq:Gpm-ineq} is
a constant multiple of $L^{mp}_0$, which is in turn estimated by
an expression in which the highest order power of $\Omega$ has
a coefficient independent of $m$ and $p$. Accordingly, we have
a bound of the form 
\begin{equation}
G^{(p)}_m(\omega,\Omega)\le   \left(c D^m \Omega^{mp} 
+\text{lower order}\right) \e^{-\beta(\omega+\Omega)^\alpha}\,.
\end{equation}

To complement the rough upper bound \eqref{eq:Gm-rough-upper}, 
we now develop a lower bound. For this purpose, we suppose that
\begin{equation}\label{eq:fhat-lower}
\hat{f}(\omega) \ge A \e^{-\beta|\omega|^\alpha}
\end{equation}
for some $A>0$. We will show that
\begin{equation}
G^{(p)}_m(\omega,\Omega) \ge A B^m \Omega^{mp}\e^{-\beta(\omega+\Omega)^\alpha}\,.
\end{equation}
where $B=A\Gamma(1+\alpha^{-1})/(2\beta)^{1/\alpha}$. 
By our assumption \eqref{eq:fhat-lower}, this statement holds for $m=0$, so suppose it is also true for some
$m\ge 0$. Using the definition of $\Xi^{(p)}$ and the fact that
all the functions appearing in the integration are positive, 
\begin{align}
G^{(p)}_{m+1}(\omega,\Omega) &\ge A^2 B^{m}\,
\int_0^\infty d\xi\, \xi^{(m+1)p}  
\,\e^{-\beta(|\Omega-\xi|^{\alpha}+|\omega+\xi|^\alpha)} \\
&\ge A^2 B^{m} \int_0^\infty d\xi\, (\Omega+\xi)^{(m+1)p}  \,\e^{-\beta(\xi^\alpha+(\omega+\Omega+\xi)^\alpha)} \\
&\ge A^2 B^{m} \Omega^{(m+1)p} \int_0^\infty d\xi\, \e^{-\beta(\xi^\alpha+(\omega+\Omega+\xi)^\alpha)} 
\end{align}
where we have first discarded the integration over $[0,\Omega]$ 
and made a change of variables $\xi\mapsto\xi+\Omega$, and
then discarded all but the highest power of $\Omega$. Now
\begin{equation}
\int_0^\infty d\xi\, \e^{-\beta(\xi^\alpha+(\sigma +\xi)^\alpha)}
\ge  \int_0^\infty d\xi\, \e^{-\beta(2\xi^\alpha+\sigma^\alpha)}= \frac{\Gamma(1+\alpha^{-1})}{(2\beta)^{1/\alpha}}   \e^{-\beta\sigma^\alpha} = \frac{B}{A}\e^{-\beta\sigma^\alpha} 
\end{equation}
for all $\sigma\ge 0$, using $(x+y)^\alpha\le x^\alpha+y^\alpha$. 
Thus 
\begin{equation}
G^{(p)}_{m+1}(\omega,\Omega) \ge AB^{m+1}\Omega^{(m+1)p}\,\e^{-\beta(\omega+\Omega)^\alpha} \,,
\end{equation}
establishing the inductive step.  

\subsection{Refined estimate of $G_m$}

We now wish to prove our statement \eqref{eq:Gm-precise} on the behavior of $G^{(p)}_m$, for which 
we require some further notation. For any $m\ge 0$, let $\Rc^{(p)}_m$ be the set of
functions $H:(0,\infty)\times(0,\infty)\to\CC$ obeying a bound of the
form 
\begin{equation}\label{eq:Rm}
\left| 
H(\omega,\Omega) 
\right|
\le \left( P(\Omega)+\frac{Q(\Omega)}{(\omega+\Omega)^{1-\alpha}}\right)
\e^{-\beta(\omega+\Omega)^\alpha}
\end{equation}
for all $\omega,\Omega>0$, where $P$ 
and $Q$ are polynomials of degrees $mp-1$ and $mp$ respectively, with nonnegative coefficients 
independent of $\omega$ and $\Omega$ (but depending on the function $H$). Here we apply the convention
 that a polynomial of degree $-1$ is identically zero. By the triangle inequality, $\Rc^{(p)}_m$
is closed under taking sums of functions, and it is also closed under
constant scalar multiples, so $\Rc^{(p)}_m$ forms a vector space
in this way. We also note, using the results of Appendix~\ref{appx:prelim}, that 
\begin{equation}\label{eq:Rc-members1}
I^r_0,~K^{q,r}_0,~L^r_0 \in \Rc^{(p)}_m \qquad \text{for $0\le r\le mp-1$, $q\ge 0$}
\end{equation}
and
\begin{equation}\label{eq:Rc-members2}
I^r_{1-\alpha},~K^{q,r}_{1-\alpha},~L^{r}_{1-\alpha}\in \Rc^{(p)}_m \qquad \text{for $1\le r\le mp$, $q\ge 0$.}
\end{equation}
It is also clear that if $H_1\in\Rc^{(p)}_m$ and $H_2(\omega,\Omega)\le H_1(\omega,\Omega)$, then $H_2\in\Rc^{(p)}_m$.

Given these definitions, \eqref{eq:Gm-precise} can be reformulated as the assertion
\begin{equation}\label{eq:Gm-pregoal}
G_{m}^{(p)} - F_{m}^{(p)}\in\Rc^{(p)}_m
\end{equation}
for all $m\ge 0$ and $p\ge 1$, where
\begin{equation}
F_{m}^{(p)}(\omega,\Omega)=
[2\pi f(0)]^m\, \Omega^{mp}\,  \hat{f}(\omega+\Omega).
\end{equation}
It will be convenient to prove an equivalent version of~\eqref{eq:Gm-pregoal}, namely that
\begin{equation}\label{eq:Gm-goal}
G^{(p)}_{m} - G^{(p)}_{m\,d}\in\Rc^{(p)}_m
\end{equation}
for all $m\ge 0$, where  
\begin{equation}
G_{m\,d}^{(p)}(\omega,\Omega) = [2\pi f(0)]^m\, \Omega^{mp}
\gamma\,  \e^{-\beta(\omega+\Omega)^\alpha}   \,.
\label{eq:Gm-dom} 
\end{equation}
The equivalence of statements \eqref{eq:Gm-pregoal} and \eqref{eq:Gm-goal} is due to the fact that $F_{m}^{(p)}- G_{m\,d}^{(p)}\in \Rc^{(p)}_m$
as a result of Eq.~\eqref{eq:f-asy}, and because $\Rc^{(p)}_m$
is closed under taking sums of functions. 

We now give an inductive proof of assertion \eqref{eq:Gm-goal}. As \eqref{eq:Gm-pregoal} holds trivially for $m=0$, the same is true of \eqref{eq:Gm-goal}. So now suppose that \eqref{eq:Gm-goal} holds for some $m\ge 0$. Recalling that $G^{(p)}_{m+1}=\Xi^{(p)}G^{(p)}_{m}$, we may write
\begin{equation}
G^{(p)}_{m+1}-G^{(p)}_{m+1\,d}=
\Xi^{(p)}\left( G^{(p)}_{m} - G^{(p)}_{m\,d}\right)
+\left(\Xi^{(p)}G^{(p)}_{m\,d}-  G^{(p)}_{m+1\,d}\right) .
\end{equation}
The inductive step is established if we show that both terms 
on the right-hand side belong to $\Rc^{(p)}_m$. 

\paragraph{First term} By the inductive hypothesis, 
$G^{(p)}_{m} - G^{(p)}_{m\,d}\in \Rc^{(p)}_m$, so 
our task is to
prove that $\Xi^{(p)}\Rc^{(p)}_m\subset \Rc^{(p)}_{m+1}$, i.e., $\Xi^{(p)}$ maps functions in $\Rc^{(p)}_{m}$ to functions in $\Rc^{(p)}_{m+1}$. To do this, it is enough to show that
$\Xi^{(p)}$ maps every function $H^r_\nu(\omega,\Omega)=\Omega^r (\omega+\Omega)^{-\nu}\,\e^{-\beta(\omega+\Omega)^\alpha}$
to an element of $\Rc^{(p)}_{m+1}$, in the cases $0\le r\le mp-1$
and $\nu=0$, or $0\le r\le mp$ and $\nu=1-\alpha$. 
Using the bound \eqref{eq:Xi-ineq}, 
\begin{equation}
|(\Xi^{(p)}H^r_\nu)(\omega,\Omega)|\le 
c\int_0^\infty d\xi\frac{\xi^{p+r}}{(\xi+\omega)^\nu}
\e^{-\beta\left(|\Omega-\xi|^\alpha+|\xi+\omega|^\alpha\right)} = 
cL^{p+r}_\nu(\omega,\Omega)
\end{equation}
and therefore our observations \eqref{eq:Rc-members1} and \eqref{eq:Rc-members2} prove that $\Xi^{(p)}H^r_\nu\in\Rc^{(p)}_{m+1}$ for the required
ranges of $r$ and $\nu$. 

\paragraph{Second term} 
We start by rewriting $\Xi^{(p)}G^{(p)}_{m\,d}$, through a change of variable as
\begin{equation}
(\Xi^{(p)}G^{(p)}_{m\,d})(\omega,\Omega)=[2\pi f(0)]^m\,\gamma\, 
  \int_{-\Omega}^\infty d\xi \, (\xi+\Omega)^{(m+1)p} \, \hat{f}(- \xi)\,
 \e^{-\beta(\omega+\Omega+\xi)^\alpha} \,.
\label{eq:G1-1}
\end{equation}
Next, using the fact that $\int_{-\infty}^\infty \hat{f}(-\xi)=2\pi f(0)$, we write the above expression as the sum of three terms
\begin{equation}
(\Xi^{(p)}G^{(p)}_{m\,d})(\omega,\Omega)= G^{(p)}_{m+1\,d}(\omega,\Omega) +R_1(\omega,\Omega)+R_2(\omega,\Omega)\,,
\end{equation}
where 
\begin{equation}
R_1(\omega,\Omega) =- G^{(p)}_{m\,d}(\omega,\Omega)\,\Omega^p \int_{-\infty}^{-\Omega} d\xi \hat{f}(- \xi) \,,
\end{equation}
and
\begin{equation}
R_2(\omega,\Omega) =[2\pi f(0)]^m\, \gamma\,\int_{-\Omega}^{\infty} d\xi \,
\hat{f}(- \xi) \, [g(\Omega+\xi) -g(\Omega)]\,,
\end{equation}
where
\begin{equation}
g(\xi) =  \xi^{(m+1)p} \, \e^{-\beta(\omega+\xi)^\alpha} \,.
\label{eq:g-def}
\end{equation}
Here $G^{(p)}_{m+1\,d}(\omega,\Omega)$ is expected to be the dominant term, and we need to show that the remainders, $R_1$ and $R_2$, lie in $\Rc^{(p)}_{m+1}$ and are therefore subdominant.  The magnitude of $R_1$ is determined from Eq.~\eqref{eq:tail},
which reveals that $R_1$ is suppressed compared to $G_{m\,d}(\omega,\Omega)$ by a factor of order $\Omega^{p+1-\alpha}\e^{-\beta \,\Omega^\alpha}$ as $\Omega\to\infty$, and of order $\Omega^p$ 
for all $\Omega>0$. Combining these estimates,  
we may bound $|R_1(\omega,\Omega)|$ by
a constant multiple of $G^{(p)}_{m\,d}(\omega,\Omega)$ and conclude that it is in $\Rc^{(p)}_{m+1}$.

To control $R_2$, we apply the mean value theorem to $g$,
obtaining
\begin{equation}
|g(\Omega+\xi)-g(\Omega)| = |\xi g'(\Omega+\eta)| \le 
|\xi|\,\e^{-\beta(\omega+\Omega+\eta)^\alpha}\left(
(m+1)p (\Omega+\eta)^{(m+1)p-1} + \frac{\alpha\beta (\Omega+\eta)^{(m+1)p}}{(\omega+\Omega+\eta)^{1-\alpha}}\right)\,,
\end{equation}
where $\eta$ lies between $0$ and $\xi$. 
(If one differentiates $g$, the two terms in parentheses appear
with a relative sign, but we have changed this to an addition for the purposes of obtaining an upper bound.) 
If $\xi<0$, we have $\xi<\eta<0$ and therefore
\begin{equation}
|g(\Omega+\xi)-g(\Omega)|   \le 
|\xi|\,\e^{-\beta(\omega+\Omega+\xi)^\alpha}\left(
(m+1)p \Omega^{(m+1)p-1} + \frac{\alpha\beta \Omega^{(m+1)p}}{(\omega+\Omega)^{1-\alpha}}\right)\,,
\end{equation}
using the fact that $(\Omega+\eta)^{(m+1)p}/(\omega+\Omega+\eta)^{1-\alpha}$ is increasing in $\eta$ on $[\xi,0]$ because $(m+1)p\ge 1-\alpha>0$. Accordingly, the contribution to $R_2$ from $[-\Omega,0]$ may be estimated by 
\begin{equation}
|R_{2\,a}| \le [2\pi f(0)]^m\, \gamma c\left((m+1)p \Omega^{(m+1)p-1}   + 
 \frac{\alpha\beta\Omega^{(m+1)p}}{(\omega+\Omega)^{1-\alpha}}\right) K^{1,0}_{0}(\omega,\Omega)\,.
\end{equation}
On the other hand, if $\xi>0$, we have $0<\eta<\xi$ and therefore
\begin{equation}
|g(\Omega+\xi)-g(\Omega)|  \le 
\xi\,\e^{-\beta(\omega+\Omega)^\alpha}\left(
(m+1)p (\Omega+\xi)^{(m+1)p-1} + \frac{\alpha\beta (\Omega+\xi)^{(m+1)p}}{(\omega+\Omega)^{1-\alpha}}\right)\,,
\end{equation}
so the contribution to $R_2$ from $[0,\infty)$ can be estimated as 
\begin{equation}
|R_{2\,b}| \le  [2\pi f(0)]^m\, \gamma c\left((m+1)p \, 
J^{1,(m+1)p-1}(\Omega) + 
\frac{\alpha\beta}{(\omega+\Omega)^{1-\alpha}}J^{1,(m+1)p}(\Omega)\right)\,\e^{-\beta(\omega+\Omega)^\alpha}\,.
\end{equation}
We see immediately that both $R_{2\,a}$ and $R_{2\,b}$ belong
to $\Rc^{(p)}_{m+1}$, which completes the proof of the inductive
step and hence establishes \eqref{eq:Gm-pregoal} for all $m\ge 0$.

\section{Asymptotics for the moments of a Lorentzian }
\label{appx:Lor}

In this appendix, we perform an asymptotic analysis of the growth rate of the moments for the special case of
a Lorentzian sampling function. In this case, the Fourier transform of the sampling function, in
$\tau = 1$ units, is given by Eq.~\eqref{eq: LorFT}. Now Eq.~\eqref{eq:Gn} may be written in the case $\ell=1$ as
\begin{equation}
G_1(\omega,\Omega) = \Omega^{p+1}  \int_0^\infty d\nu \, \nu^p \,
 \exp\left\{-\Omega \left[\left(\frac{\omega}{\Omega}+\nu \right) +|1-\nu|  \right]\right\}\,. 
 \label{eq:G1-L}
\end{equation}
We define $G_{1a}(\omega,\Omega)$ as the contribution from the interval $1 \leq \nu < \infty$,
and change the integration variable to $s =\nu -1$, leading to
\begin{equation}
G_{1a}(\omega,\Omega) = \Omega^{p+1} \, \int_0^\infty ds \, (1+s)^p \, {\rm e}^{-\Omega(2s +1 +\omega/\Omega)}
\approx \frac{1}{2}\, \Omega^p\,  {\rm e}^{-(\omega + \Omega)} \,.
\end{equation}
In the last step, we used $(1+s)^p \approx 1$, as the region where $s \ll 1$ dominates for large $\Omega$.
If we define $G_{1b}(\omega,\Omega)$ as the contribution to Eq.~\eqref{eq:G1-L} from the interval $0 \leq \nu \leq 1$,
we may use $\omega/\Omega +\nu + |1-\nu| = \omega/\Omega + 1$ to write an exact result:
\begin{equation}
G_{1b}(\omega,\Omega) = \frac{ \Omega^{p+1}}{p+1}\;  {\rm e}^{-(\omega + \Omega)} \,.
\end{equation}
Now we have the asymptotic form
\begin{equation}
G_{1}(\omega,\Omega) = G_{1a}(\omega,\Omega) + G_{1b}(\omega,\Omega) \approx  
 \frac{ \Omega^{p+1}}{p+1}\;  {\rm e}^{-(\omega + \Omega)} \, \left[ 1 + \frac{p+1}{ 2 \Omega} + O(\Omega^{-2}) \right]\,.
 \label{eq:G1asym}
\end{equation}
This result can be compared with one obtained from Eq.~(A4) of Ref.~\cite{FFR12}, which leads to
\begin{equation}
G_{1}(\omega,\Omega) = \frac{p!}{2^{p+1}}\, {\rm e}^{-(\omega + \Omega)} \,
\sum_{r=0}^{r=p+1} \frac{(2 \Omega)^r}{r!}\,, 
\end{equation}
for which the $r=p+1$ and $r=p$ terms reproduce Eq.~\eqref{eq:G1asym} of the present
paper.

The recurrence relation,   Eq.~\eqref{eq:Gn}, may be expressed for the Lorentzian case as
\begin{equation}
G_{m+1}^{(p)}(\omega,\Omega) = 
\int_0^\infty d\nu \, \nu^p \, G_{m}^{(p)}(\omega,\nu)\, {\rm e}^{-|\nu-\Omega|} \,.
\label{eq:Gm-L}
\end{equation}
In this case, we replace Eq.~\eqref{eq:Gm-ans} by the ansatz
\begin{equation}
G_{m}^{(p)}(\omega,\Omega) \approx \Omega^{m(p+1)}\, {\rm e}^{-(\omega +\Omega)}\,
\left[a_m +\frac{b_m}{2 \Omega} O(\Omega^{-2}) \right]\,.
\label{eq:Gm-ans-L}
\end{equation}
where
\begin{equation}
a_1 = \frac{1}{p+1} \; {\rm and} \; b_1 =1\,.
\label{eq:init}
\end{equation}
Next we use Eq.~\eqref{eq:Gm-ans-L} in Eq.~\eqref{eq:Gm-L}  to write
\begin{eqnarray}
& &G_{m+1}^{(p)}(\omega,\Omega) = a_m\, G_{1}^{(m+1)(p+1)-1}(\omega,\Omega) 
+\frac{b_m}{2}\,  G_{1}^{(m+1)(p+1)-2}(\omega,\Omega) \nonumber \\
&\sim&
\Omega^{(m+1)(p+1)} {\rm e}^{-(\omega +\Omega)}
\left[ \frac{a_m}{(m+1)(p+1)} + \frac{1}{2 \Omega} \left( a_m + \frac{b_m}{(m+1)(p+1) -1} \right)\right] \,,
\end{eqnarray}
which leads to the recurrence relations
\begin{equation}
a_{m+1} = \frac{a_m}{(m+1)(p+1)}\,,
\end{equation} 
and
\begin{equation}
b_{m+1} = a_m + \frac{b_m}{ (m+1)(p+1) -1} \,.
\end{equation}
These relations may be combined with Eq.~\eqref{eq:init}  to find
\begin{equation}
a_m = \frac{1}{m! \, (p+1)^m} \,,
\end{equation}
and
\begin{equation}
b_m = \frac{(m+1)(p+1) -1}{(m-1)! \, (p+1)^{m-1}\, (2p+1)} \,.
\end{equation}

We may now combine Eqs.~\eqref{eq:M} and \eqref{eq:Gm-ans-L} to write an
asymptotic form for the even moments as
\begin{eqnarray}
M_{2m} &=& C_{2m}   \int_0^\infty d\omega d\Omega \, (\omega \Omega)^p\, 
[G_{m-1}(\omega, \Omega)]^2 \nonumber \\
&\sim&  C_{2m}   \int_0^\infty d\omega d\Omega \, (\omega \Omega)^p\, 
\Omega^{2(m-1)(p+1)}\, {\rm e}^{-2(\omega +\Omega)}\, 
\left(a_{m-1}^2 +\frac{a_m\,b_{m-1}}{\Omega}\right) \nonumber \\
&=&  C_{2m}\ \frac{p!}{2^{p+1}} \, \frac{[(2m-1)(p+1) -1]!}{2^{(2m-1)(p+1)}}\;
\left(a_{m-1}^2  + \frac{2\, a_{m-1}\, b_{m-1}}{(2m-1)(p+1) -1}\right) \nonumber \\
&=& C_{2m} \, \frac{p!\, [(2m-1)(p+1) -1]!}{2^{2m(p+1)}\, [(m-1)!]^2\, (p+1)^{2(m-1)}} \; X \,,
\label{eq:MmL}
\end{eqnarray}
where
\begin{equation}
X = 1+ \frac{2 (m-1)(p+1)[m(p+1) -1]}{(2p+1) [(2m-1)((p+1)-1]}\,.
\end{equation}
Note that if $m \gg 1$,
\begin{equation}
X \sim 1 + \frac{m(p+1)}{2p+1} \,,
\end{equation}
or
\begin{equation}
X \sim \left[1 + \frac{(p+1)}{2(2p+1)} \right]^{2m} \,. 
\end{equation}

Stirling's formula,
\begin{equation}
n! \sim \sqrt{2\pi}\, n^{n+1/2}\, {\rm e}^{-n}\qquad n \ll 1   \,,
\end{equation}
may be used to write 
\begin{eqnarray}
(A m -B)! &\sim& \sqrt{2\pi}\,(A m)^{Am-B +1/2}\, 
\left(1-\frac{B}{A m}\right)^{B\, (A m)/B -B +1/2}\,   {\rm e}^{-(Am-B)} \nonumber \\
&\sim& \sqrt{2\pi}\, \left(\frac{Am}{e}\right)^{Am-B +1/2}\,  {\rm e}^{-B +1/2} \,.
\label{eq:AB}
\end{eqnarray}
In the last step, we used,
\begin{equation}
\lim_{x \rightarrow \infty} \left(1- \frac{1}{x}\right)^x = \frac{1}{e} \,,
\end{equation}
and 
\begin{equation}
\left(1-\frac{B}{A m}\right)^{-B +1/2} \sim 1 \,.
\end{equation}
We may show that
\begin{equation}
 \frac{[(2m-1)(p+1) +1]!}{ [(m-1)!]^2} \sim 
 \frac{1}{4\pi}\, 2^{2m}\, (p+1)^{2m(p+1) -p-3/2}\, p^{p(1-2m) +1/2}\, [2p m -(p+1)]! \,,
\end{equation}
by application of Eq.~\eqref{eq:AB} to the factorials on both sides of this relation.

This allows us to rewrite Eq.~\eqref{eq:MmL} as
\begin{equation}
M_{2m} \sim \frac{C_{2m}}{4\pi}\, (p+1)! \, \left(\frac{p}{p+1}\right)^{p+1/2} \,
\left[\frac{(p+1)^p}{2^p\,p^p}\right]^{2m} \, X\, [2p m -(p+1)]! \,.
\end{equation}
For the case $p=3$, this leads to $M_n \propto (3n-4)!$, which is the rate of growth
found in Ref.~\cite{FFR12}. This rate of growth in turn leads to the asymptotic behavior
of the probability distribution given in Eq.~\eqref{eq:Lor}.

\end{document}